\documentclass[prd,reprint, nofootinbib,amsmath,amssymb, aps, floatfix]{revtex4-1}
\usepackage{dcolumn}% Align table columns on decimal point
\usepackage{bm}% bold math
\usepackage{amssymb,amsmath,graphicx}
\usepackage[linktoc=page]{hyperref}
\usepackage{epsfig}
\usepackage{amsfonts}
\usepackage{tikz}
\usetikzlibrary{decorations.pathmorphing}
\usepackage{graphicx}
\usepackage{float}
\begin{document}
%\normalem
\title{Dynamics and Observer-Dependence of Holographic Screens}
\author{Raphael Bousso and Mudassir Moosa}%
 \email{bousso@lbl.gov, mudassir.moosa@berkeley.edu}
\affiliation{ Center for Theoretical Physics and Department of Physics\\
University of California, Berkeley, CA 94720, USA 
}%
\affiliation{Lawrence Berkeley National Laboratory, Berkeley, CA 94720, USA}
\begin{abstract}
We study the evolution of holographic screens, both generally and in explicit examples, including cosmology and gravitational collapse. A screen $H$ consists of a one-parameter sequence of maximal surfaces called leaves. Its causal structure is nonrelativistic. Each leaf can store all of the quantum information on a corresponding null slice holographically, at no more than one bit per Planck area. Therefore, we expect the screen geometry to reflect certain coarse-grained quantities in the quantum gravity theory. 

In a given spacetime, there are many different screens, which are naturally associated to different observers. We find that this ambiguity corresponds precisely to the free choice of a single function on $H$. We also consider the background-free construction of $H$, where the spacetime is not given. The evolution equations then constrain aspects of the full spacetime and the screen's embedding in it.
\end{abstract}
%\pacs{}
\maketitle
\tableofcontents

\section{Introduction}
\label{intro}

In the search for a quantum theory of gravity in general spacetimes, the study of holographic screens~\cite{CEB2} has recently led to interesting new results. An area theorem was proven for past and future holographic screens in any spacetime satisfying the null curvature condition~\cite{BouEng15a,BouEng15b}. The semiclassical extension of this theorem led to the first rigorous formulation of a universal Generalized Second Law~\cite{BouEng15c}, applicable in cosmology and other highly dynamical spacetimes. In the present paper, we will study the classical evolution of holographic screens in more detail.

\begin{figure*}[t]
%\centering 
\begin{tikzpicture}
\draw [orange,dashed](0,-3) --(4,1);
	\draw [orange,dashed](-.75,-2.25) --(3.25,1.75);
	\draw [orange,dashed](-1.5,-1.5) --(2.5,2.5);
	\draw [orange,dashed](-2.25,-0.75) --(1.75,3.25);
	\draw [orange,dashed](-3,0) --(1,4);
	\draw [green,thick](-.6,0.1) --(-.25,-0.25);
	\draw [green,thick,->](-1,0.5) --(-.6,0.1);
	\draw [orange,thick](.2,0.2) --(.5,0.5);
	\draw [orange,thick,->](-.25,-0.25) --(.2,0.2);
	%%%
	\draw [green,thick](.95,0.05) --(1.25,-0.25);
	\draw [green,thick,->](.5,0.5) --(.95,0.05);
	\draw [orange,thick](1.75,0.25) --(2.3,0.8);
	\draw [orange,thick,->](1.25,-0.25) --(1.75,0.25);
	\node at (-2.0,2.50) {$H$};
	\draw [blue!50!white,thick](-2.5,2.5) to (-1.75,1.25) to [out=290,in=150] (-1,0.5) to [out=330,in=200] (.5,0.5) to [out=20,in=175] (2.3,0.8) to [out=0,in=145] (3.4,0.4) to (4.6,-0.6);
	\fill [blue](-1.75,1.25) circle (.1cm);
	\fill [blue](-1.0,0.5) circle (.1cm);
	\fill [blue](0.5,0.5) circle (.1cm);
	\fill [blue](2.3,0.8) circle (.1cm);
	\fill [blue](3.4,0.4) circle (.1cm);
	\fill [red](-.25,-.25) circle (.1cm);
	\fill [red](1.25,-.25) circle (.1cm);
	%	\draw [fill=gray!60!white](0,-9) to [out=70,in=270] (2,5) --(0,5) --(0,-9);
	\node at (-1.5,0.45) {\scriptsize$\sigma(R)$};
	\node at (0.25,0.75) {\scriptsize$\sigma(R+dR)$};
	\node at (2.2,1.0) {\scriptsize$\sigma(R+2dR)$};
	\node at (-0.70,-0.52) {\scriptsize$\bar{\sigma}(R+dR)$};
	\node at (1.4,-.55) {\scriptsize$\bar{\sigma}(R+2dR)$};
	\node at (2.05,3.45) {$N(R)$};
	\node at (2.8,2.7) {$N(R+dR)$};
	\node at (3.45,1.95) {$N(R+2dR)$};
	\node at (0.70,-.10) {$\alpha l^{a}$};
	\node at (2.05,0.15) {$\beta k^{a}$};
%%%%%%%%%%%%%%%%%%%%%%%%%%%%%%%%%%%%%%%%%%%%%%%%
%\node at (0,0) {\includegraphics[scale=0.4]{figure-ab-2.jpg}};
%\node at (3.35,.89) {\small{$\bar{\sigma}(R_{2}+dR)$}};
%\node at (3.1,-1.5) {\small{$\bar{\sigma}(R_{1}+dR)$}};
%\node at (1.85,2) {\small{$\sigma(R_{3})$}};
%\node at (1.6,-0.4) {\small{$\sigma(R_{2})$}};
%\node at (1.35,-2.8) {\small{$\sigma(R_{1})$}};
%\draw [->](-1.05,-2.8) --(-2.15,-1.7);
%\node at (-1.75,-2.45) {\small{$\alpha l^{a}$}};
%\draw [->](-2.15,-1.3) --(-1.30,-0.45);
%\node at (-1.9,-0.8) {\small{$\beta k^{a}$}};
%\draw [->](-1.30,-0.35) --(-2.40,0.75);
%\node at (-2.1,0.1) {\small{$\alpha l^{a}$}};
%\draw [->](-2.40,1.1) --(-1.45,2.05);
%\node at (-2.1,1.65) {\small{$\beta k^{a}$}};
%\node at (-1.7,2.85) {H};
%\node at (0.1,-0.9) {\scriptsize{N($R_{1}$+dR)}};
%\node at (0.1,1.4) {\scriptsize{N($R_{2}$+dR)}};
%%%%%%%%%%%%%%%%%%%%%%%%%%%%%%%%%%%%%%%%%%%%%%%%%
%%%%%%%%%%%%%%%%%%%%%%%%%%%%%%%%%%%%%%%%%%%%%%%%%
\node at (9,0) {\includegraphics[scale=0.4]{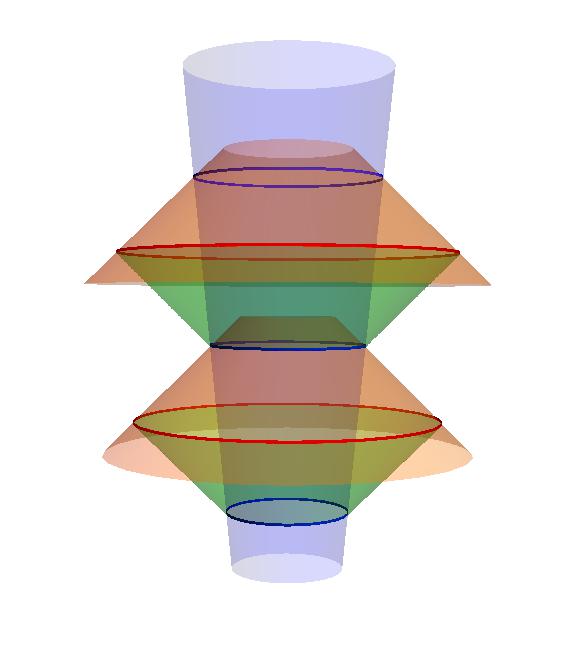}};
\node at (12.35,.96) {\small{$\bar{\sigma}(R+2dR)$}};
\node at (12.05,-1.4) {\small{$\bar{\sigma}(R+dR)$}};
\node at (11.27,2.05) {\small{$\sigma(R+2dR)$}};
\node at (11.0,-0.4) {\small{$\sigma(R+dR)$}};
\node at (10.32,-2.8) {\small{$\sigma(R)$}};
\draw [->](7.95,-2.8) --(6.85,-1.7);
\node at (7.25,-2.45) {\small{$\alpha l^{a}$}};
\draw [->](6.85,-1.25) --(7.70,-0.40);
\node at (7.05,-0.8) {\small{$\beta k^{a}$}};
\draw [->](7.70,-0.35) --(6.60,0.75);
\node at (6.9,0.1) {\small{$\alpha l^{a}$}};
\draw [->](6.60,1.1) --(7.55,2.05);
\node at (6.9,1.65) {\small{$\beta k^{a}$}};
\node at (7.3,2.85) {H};
\node at (6.05,-2.25) {\footnotesize{N($R+dR$)}};
\node at (5.65,0.3) {\footnotesize{N($R+2dR$)}};
\end{tikzpicture}
\caption{{\em Left:} a future holographic screen, $H$ (blue line). Points represent topological spheres. The dashed lines are 2+1 dimensional null slices $N$ orthogonal to the leaves $\sigma$ of $H$ (blue dots), along the direction $k^a$. $H$ can be constructed leaf by leaf, using a ``zig-zag'' procedure. First, deform the leaf $\sigma(R)$ along the other orthogonal null vector, $l$, by an infinitesimal step $\alpha(R,\vartheta,\varphi) l^a$ (green downward arrow). The function $\alpha<0$ can be chosen arbitrarily; it reflects a kind of observer-dependence of the holographic screen. Thus one obtains a new surface $\bar\sigma(R+dR)$ (red), and from it, a new null slice $N(R+dR)$ orthogonal to $\bar\sigma$. The next leaf $\sigma(R+dR)$ is the surface of maximal area on $N(R+dR)$, at some infinitesimal distance $\beta k^a$ along $N$ from $\bar\sigma$ (orange arrow). {\em Right:} a past holographic screen (same color coding).  In this case $\alpha>0$; the area of the leaves grows towards the future. We show the same construction, with only one spatial direction suppressed to offer a different visualization. The leaves $\sigma(R)$ are by definition the maximal area cross-sections of $N(R)$, despite what the figure shows.}
\label{figure-ab}
\end{figure*}
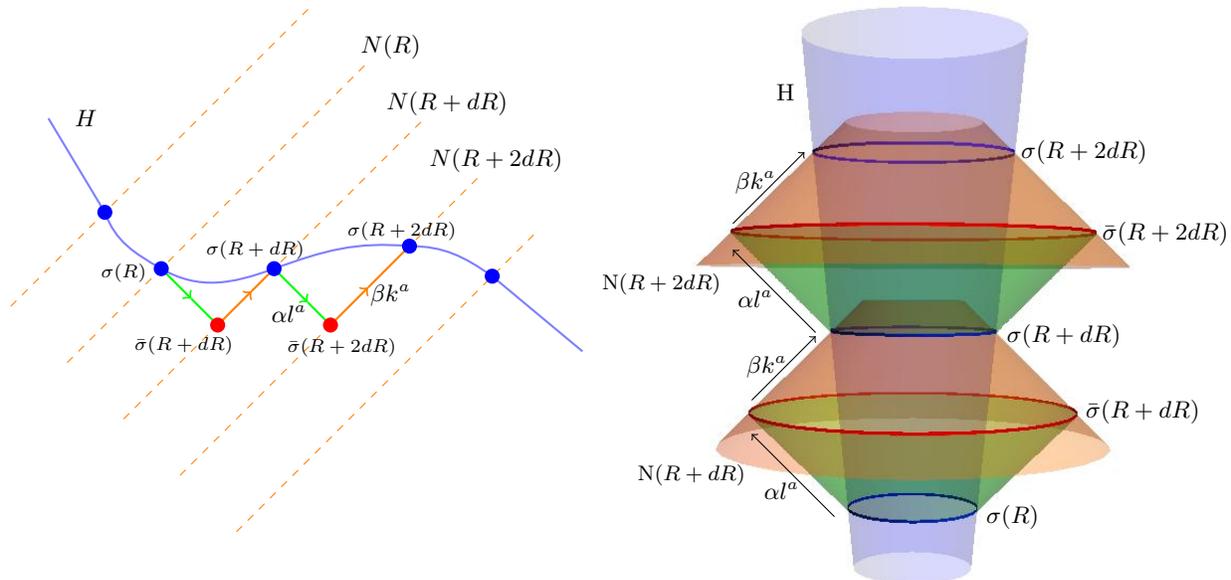
 
A {\em holographic screen} $H$ can be associated to a null foliation of a spacetime $M$, i.e., a foliation of $M$ into 2+1 dimensional hypersurfaces $N(R)$, each with two spatial and one light-like direction. (See Fig.~\ref{figure-ab}.) The screen consists of a sequence of two-dimensional surfaces $\sigma(R)$ called {\em leaves}. Each leaf is the spatial cross-section of largest area on the corresponding slice $N(R)$. A holographic screen is called {\em future} ({\em past}) if the area of each leaf is decreasing (increasing) in the opposite light-like direction, i.e., if every $\sigma(R)$ is marginally trapped (anti-trapped). Future screens appear inside black holes or near a big crunch. Past holographic screens exist in an expanding universe, for example in ours.

The covariant entropy bound (Bousso bound)~\cite{CEB1,FMW} implies that all of the information about the quantum state on each null slice $N(R)$ can be stored on the corresponding leaf $\sigma(R)$, at a density of no more than one bit per Planck area. This suggests that the holographic principle~\cite{Tho93,Sus95,FisSus98} applies in all spacetimes. (Several precise semiclassical versions of this conjecture have recently been formulated, and in some cases, proven rigorously~\cite{BouCas14a,BouCas14b,BouFis15a,BouFis15b,KoeLei15}.) The holographic relation between quantum information and geometry substantially involves both $G_N$ and $\hbar$, Newton's and Planck's constants. Its origin can only lie in a quantum theory of gravity, so one expects the structure of holographic screens to reflect aspects of the underlying theory.

In spacetimes with conformal boundaries, all or parts of the screen lie on it~\cite{CEB2,RMP}. For example, in asymptotically Anti-de Sitter spacetimes, the screen is located on the conformal boundary at spatial infinity. This is consistent with the AdS/CFT correspondence providing a full quantum description. 

It is therefore of interest to study holographic screens in more realistic spacetimes, where quantum gravity remains a mystery. (An interesting recent approach explores a generalization of the stationary-surface conjecture~\cite{RyuTak06,HubRan07} for computing entanglement entropy~\cite{SanWei16a,SanWei16b,NomSal16}.) In particular, it is important to understand the dynamics of holographic screens in cosmology and in the collapsing regions inside of black holes.

Future holographic screens have already been studied in some detail under the guise of ``dynamical horizons''~\cite{AshKri02} or ``future outer-trapped horizons''~\cite{Hayward-TH}, as interesting candidates for quasi-local boundaries of black holes. Strictly, the latter objects are more restrictive: dynamical horizons correspond only to the {\em spacelike} portions of future holographic screens. For the purposes of proving an area theorem, the restriction to spacelike portions is significant: the area theorem is trivial for dynamical horizons, but highly nontrivial for future holographic screens. This is because without the spacelike assumption, the area theorem relies on a global property that is hard to prove: the (unique) foliation of a given screen $H$ into leaves $\sigma(R)$ uniquely defines a foliation of a (portion of) the spacetime $M$ into null slices $N(R)$.

Here we will be interested in studying the evolution of local quantities, the metric and extrinsic curvature of the leaves. For this purpose, the spacelike assumption yields no significant simplification. In fact, a number of authors have studied the local evolution problem for dynamical horizons~\cite{AshKri02,AshKri03,AshKri04,Gourgoulhon-TH,Gourgoulhon-DH,Gourgoulhon-DH-1}, and some noted that the spacelike assumption is not required for the validity of the evolution equations. For completeness, we offer a simplified derivation of the evolution equations in the Appendix~\ref{der_screen_eqs}. In the main text, our focus will be on their interpretation. In particular, we will emphasize the role of a gauge choice which corresponds geometrically to a choice of null foliation, and which has a natural interpretation as reflecting a choice of observer.

In Sec.~\ref{sec-hs}, we establish conventions, and we define the screen variables: local geometric quantities that can be associated to a holographic screen. They include the metric and null extrinsic curvatures of the leaves, a tangent vector field to the leaves describing the relative evolution of the two null normals, and a tangent vector field normal to the leaves describing the ``slope'' and ``rate'' of the screen's progress through the spacetime it is embedded in. 
We also identify one ``global'' and one ``gauge'' transformation, which leave the screen invariant but act nontrivially on some of the above variables.

In Sec.~\ref{sec-dod}, we present the evolution equations for the screen variables. We then analyze them from three different perspectives. First, in Sec.~\ref{sec-dod1}, we regard both the spacetime $M$ and the screen $H$ as given. This viewpoint has been examined previously, and it has led to suggestions that the screen evolution can be interpreted as fluid dynamics. We identify a number of problems with this interpretation. 

Next, in Sec.~\ref{sec-dod2}, we regard the spacetime $M$ as given but consider the evolution equations as a tool for constructing $H$. We find that the equations are underdetermined by one function $\alpha$ on $H$. We show that this function corresponds precisely to the ambiguity in choosing a null slicing; see Fig.~\ref{figure-ab}. More precisely, given a partially constructed screen up to some leaf $\sigma(R)$, we show that $\alpha$ can be regarded as a lapse function that describes how much the infinitesimal step $R$ advances the slicing away from each point on $\sigma(R)$. This defines a new null slice $N(R+dR)$ and ultimately, a new leaf $\sigma(R+dR)$, in an $\alpha$-dependent way. 

We can regard $\alpha$ as encoding a kind of generalized observer-dependence of the screen, in the following sense. Consider a worldline, and consider the future light-cone from each point on the worldline. If the worldline is in a collapsing region (e.g., inside a black hole), then there will be a cross-section of maximum area on this light-cone: a marginally trapped surface. The sequence of such surfaces defined by the above construction yields a holographic screen, $H$. 

Now consider a different observer, whose worldline coincides in some interval with the previous one, but then departs from it. The above construction still works, and in the region where the worldlines agree it, it will yield the same leaves. Therefore the holographic screens will also agree on those leaves. But the leaves constructed from the light-cones of points where the worldlines do not agree will differ. Therefore there is no unique future evolution for a holographic screen, even if we are given part of the screen and the entire spacetime $M$.

This mathematical description of the observer-dependence of holographic screens, as a choice of the function $\alpha$, is the central result of this paper. It would be nice to explore this further. For example, infinitesimally nearby screens encode nearly the same subset of $M$. The transformation relating them may correspond either to a change of variables in the underlying theory, or to a change of the prescription for reconstructing spacetime from those variables.

Finally, in Sec.~\ref{sec-dod3}, we consider the evolution equations from a ``background-free'' perspective, where neither $M$ nor $H$ are given. In this case, we can regard the screen variables as given. What was previously regarded as their evolution equations now determines aspects of the spacetime $M$, and of how $H$ is embedded in $M$. However, from this viewpoint the equations are highly underdetermined. This is not surprising, since the screen variables can at most represent a coarse grained subset of the information in the underlying quantum gravity theory.

In Sec.~\ref{sec-examples}, we illustrate our general analysis with some examples. We construct screens explicitly for black holes and for cosmological solutions, and we compute the screen variables. In particular, we construct two different screens for the same cosmology, only one of which is spherically symmetric. This illustrates the observer-dependence associated with a choice of different worldlines and null slicings.

\paragraph*{Relation to Other Work}

Our analysis builds on earlier studies of dynamical horizons and future outer-trapped horizons, such as Refs.~\cite{Hayward-TH,AshKri02,AshKri03,Hayward-law-1,Hayward-law-2,Booth-Hamiltonian}. In many of these works, an analogue of the first law of black hole thermodynamics was sought. (The second law holds trivially for dynamical horizons.) However, it is not clear that physically meaningful intrinsic and extrinsic variables, such as total energy and temperature, can be uniquely defined. We do not pursue this direction here, though we note in Sec.~\ref{sec-examples} that a certain local geometric quantity $\kappa$ limits to the usual surface gravity of an event horizon, in all examples where a sensible comparison can be made.

Here, we focus on local parameters that arise naturally from the geometry of holographic screens. In Sec.~\ref{sec-dod} we take as our starting point the evolution equations of Gourgoulhon and Jaramillo~\cite{Gourgoulhon-TH,Gourgoulhon-DH,Gourgoulhon-DH-1}. (For completeness, their derivation is given in the Appendix~\ref{der_screen_eqs}.) In Sec.~\ref{sec-examples}, we make use of the work of Booth {\em et al.}~\cite{Booth-examples}, who explicitly constructed dynamical horizons for spherical dust collapse.

\section{Kinematics of Holographic Screens}
\label{sec-hs}

A {\em future (past) holographic screen}, $H$, is a hypersurface (not necessarily of definite signature) that is foliated by marginally trapped (anti-trapped) codimension-2 spatial surfaces called {\em leaves}. For simplicity we will take spacetime to have four dimensions in what follows, and we consider future screens unless otherwise noted; but all results are easily generalized. By a {\em surface} we shall mean a smooth two-dimensional achronal surface. We will consider only regular screens, which satisfy a set of further mild technical conditions~\cite{BouEng15b} such as the generic condition, Eq.~(\ref{eq-generic}) below.  In this section, we will discuss the kinematic structure underlying holographic screens and establish a number of conventions.

%We will encounter two key ambiguities which pervade the literature on dynamical horizons and FOTHs: the parametrization of the leaves $\sigma(R)$, and the normalization of their normal vector tangent to $H$, $h$. We will find that the recent area theorem~\cite{BouEng15a} and its proof~\cite{BouEng15b} suggest preferred choices that remove this freedom.

\subsection{Tangent and Normal Vectors}
\label{sec-vectors}

In a Lorentzian manifold, every two-dimensional spatial surface has two future-directed orthogonal null vector fields, $k^{a}$ and $l^{a}$. It is convenient to choose their normalization such that
\begin{equation}
k^{a}l_{a}=-1~.
\label{eq-kl}
\end{equation}
This allows for arbitrary rescalings $l\to\gamma l$, $k\to \gamma^{-1} k$, where $\gamma$ is an arbitrary positive function on the screen $H$. We show below that this gives rise to a $U(1)$ gauge symmetry.
%This leaves one parameter, the normalization of $k$, which we will fix below. 

A surface is marginally trapped if
\begin{equation}
\theta^{(k)} = 0~,~~ \theta^{(l)}<0~.
\end{equation}
By the above definition, a future holographic screen can be thought of as a one-parameter sequence of such surfaces, its leaves $\sigma(R)$. In principle, any parameter can be used. For example, the existence of an area theorem for holographic screens~\cite{BouEng15a,BouEng15b} makes it possible to choose $R$ to be a monotonic function of the area of the leaves.
%We shall take $R$ to be the ``area radius,''
%\begin{equation}
%R \equiv \sqrt{\frac{A}{4\pi}}~.
%\label{eq-RsA}
%\end{equation}

Next we wish to define a vector field $h$ which is tangent to $H$ and normal to each leaf $\sigma(R)$. The latter condition implies that
\begin{equation}
h^{a} = \alpha l^{a} + \beta k^{a}~.
\label{eq-hbeta}
\end{equation}
%and the former fixes the ratio $\beta/\alpha$ given a choice of $l$. 
A key intermediate result in the proof of the area theorem~\cite{BouEng15b} is that $\alpha<0$ everywhere on $H$ (in our convention where $l$ is future-directed). That is, the evolution of leaves of a future holographic screen is towards the past or the spatial exterior.

The parameter $\beta$ corresponds to the ``slope'' of the holographic screen. By Eq.~(\ref{eq-hbeta}), the screen is past-directed if $\beta<0$ and spatially outward-directed if $\beta>0$. The generic condition of Ref.~\cite{BouEng15b} prevents $h$ from becoming collinear with $k$, so $\beta$ is always finite. However, $\beta$ has no upper bound. In the limit as $\beta\to\infty$, $H$ approaches an isolated horizon. For example, $H$ can approach the event horizon of a black hole from the inside. 

Because $\beta$ can have any sign, $H$ need not be of definite signature. Thus we cannot require that $h$ have unit norm:
\begin{equation}
h^a h_a = -2\alpha\beta~.
\label{eq-hnorm}
\end{equation}
Instead, we normalize $h$ by requiring that
\begin{equation}
h(R) = h^a (dR)_a = 1~,
\label{eq-hR}
\end{equation}
where $R$ is the (arbitrary) foliation parameter. 
We also define a vector normal to $H$ and to every leaf:
\begin{equation}
n^{a} = -\alpha l^{a} + \beta k^{a} ~,
\label{form_n}
\end{equation}
which satisfies $h^a n_a=0$ and $n^a n_a=2\alpha\beta$.  

There are two ways to think about this normalization, corresponding to different perspectives on screen evolution. In one viewpoint, we consider a given screen $H$ in a given spacetime $M$. Then it is natural to choose a foliation parameter $R$, which fixes the product $\alpha\beta$ via the above two equations. The ratio $\beta/\alpha$ is fixed by the slope of the screen's embedding in $M$. 

Alternatively, we may consider only the spacetime $M$ as given, and consider it our task to construct the screen $H$. In this case the screen will not be unique. Even if some portion of the screen is known (as a set of leaves associated with a finite range of $R$), this does not determine the remainder of the screen. We shall see that the ambiguity is precisely associated with a choice of a negative function $\alpha$ on $H$ (at fixed choice of $l$). This corresponds to a choice of null foliation of $M$, or physically, to a choice of observer associated with the screen. We will later identify a constraint equation that determines $\beta$ as a function of $\alpha$ and other data, Eq.~(\ref{eq-bced}) below. The parameter $R$ is then determined by Eq.~(\ref{eq-hR}).

% The only remaining ambiguity are the rescalings of $l$. To fix this, we make use of a This allows us to choose
% \begin{equation}
% \alpha=-1
% \label{eq-alpha}
% \end{equation}
% everywhere on $H$, without violating the condition that $l$ be future-directed. 

% Given a choice of screen parameter $R$, Eqs.~(\ref{eq-kl}), (\ref{eq-hR}), and (\ref{eq-alpha})  completely determines $R$, $k$, $l$, and $h$. We now have
% \begin{equation}
% h^{a} = - l^{a} + \beta k^{a}~,~~h^a h_a = 2\beta~.
% \label{eq-hbeta}
% \end{equation}

The induced metric on the screen $H$ is not always well-defined:
\begin{equation}
\gamma_{ab} = g_{ab} - \frac{1}{2\alpha\beta}n_{a}n_{b}~;
\end{equation}
This is ill-defined on null portions of $H$, i.e., when $\beta$ vanishes; and it changes signature when $\beta$ changes sign. But we will not need this metric below. 

By contrast, the induced spatial metric on a leaf $\sigma(R)$ is always well defined:
\begin{equation}
q_{ab} = g_{ab} + k_{a}l_{b} + l_{a}k_{b}~.% =\mbox{ }g_{ab} + \frac{1}{2\beta}n_{a}n_{b} - \frac{1}{2\beta}h_{a}h_{b} 
\label{metric_leaf}
\end{equation}

\subsection{Extrinsic Curvature and Acceleration}
\label{sec-ext}

We are interested in the extrinsic curvature of the leaves $\sigma(r)$ in the spacetime, rather than the extrinsic curvature of the screen $H$. Since the leaves are of codimension 2, the full extrinsic curvature data consists of the following objects: the null extrinsic curvatures in the $k$ and $l$ directions, respectively; and the so-called Weingarten map, which measures how the null normals vary with respect to each other. 

The null extrinsic curvatures are defined by
\begin{eqnarray} 
B^{(k)}_{ab} & = & q_a^c q_b^d \nabla_c k_d ~,\\
B^{(l)}_{ab} & = & q_a^c q_b^d \nabla_c l_d ~.
\end{eqnarray}
The expansion and shear are given by 
\begin{eqnarray} 
\theta^{(k)} & = & B^{(k)}_{ab} q^{ab}\label{eq-expansion}\\
\sigma^{(k)}_{ab} & = & B^{(k)}_{(ab)}-\frac{1}{2} \theta^{(k)} q_{ab}~,\label{eq-shear}
\end{eqnarray}
and similarly for $l$. We recall that by definition of a future holographic screen, $\theta^{(k)}=0$ and $\theta^{(l)}<0$.

Analagously one can define extrinsic curvature, expansion, and shear for any vector field orthogonal to $\sigma$, such as $h^a$ or $n^a$. Since the definitions are linear, Eq.~(\ref{eq-hbeta}) implies, e.g., 
\begin{eqnarray} 
\theta^{(h)} & = & \alpha\theta^{(l)}+ \beta \theta^{(k)} = \alpha\theta^{(l)}~,\label{eq-thetahl}\\
\theta^{(n)} & = & -\alpha\theta^{(l)}+ \beta \theta^{(k)} = -\alpha\theta^{(l)}~. \label{eq-thetanl}
\end{eqnarray}

From the one-form $-l_{b}\nabla_{a}k^{b}$, one can construct the {\em normal one-form} by projection along the leaf,
\begin{equation} 
\Omega_a \equiv q_{a}^{\mbox{ }c} (-l_{b}\nabla_{c}k^{b}) ~, \label{eq-omega-one-form}
\end{equation} 
and the {\em acceleration} $\tilde{\kappa}$ by projection along the evolution vector field,
\begin{equation} 
\tilde\kappa\equiv h^c(-l_{b}\nabla_{c}k^{b}) ~.
\label{eq-kappa}
\end{equation} 
This quantity is called ``surface gravity'' in Refs.~\cite{Gourgoulhon-TH,Gourgoulhon-DH,Gourgoulhon-DH-1} and is denoted $\kappa$ there. We will reserve that term and notation for a different, closely related quantity defined in Eq.~(\ref{eq-kktb}) below, because we find that it better matches the surface gravity of event horizons.

It is easy to see that the following expressions are equivalent to Eq.~(\ref{eq-kappa}): $\tilde\kappa = k_b h^a \nabla_a l^b = h_b h^a \nabla_a k^b = -l_b h^a\nabla_a(h^b/\beta)$. Yet another equivalent expression for $\tilde{\kappa}$ can be given by extending the null vector fields $k$ and $l$ into a neighborhood of $H$ (which was not needed above), according to the following prescription: $l$ is parallel transported along itself, and $k$ is parallel transported but rescaled so as to satisfy  $k^a l_a=-1$ everywhere. With this choice, one finds
\begin{equation}
k^{a}\nabla_{a}k^{b} =\kappa k^{b} ~,
\label{k_off_H}
\end{equation}
where
\begin{equation}
\kappa \equiv \frac{\tilde\kappa}{\beta}~.
\label{eq-kktb}
\end{equation}
At points where $\beta=0$, the above prescription fails to extend $l$ into an open neighborhood of such points, leading to a divergence.

Notably, Eq.~(\ref{k_off_H}) takes the same form as the definition of the surface gravity of a Killing horizon. However, the acceleration $\kappa$ is not invariant under certain allowed rescalings of $k$, which we will discuss shortly. For Killing horizons, there is a similar ambiguity, which would also rescale the surface gravity. But in some cases (e.g.\ asymptotically flat spacetimes), a preferred normalization of the Killing vector field $k_{\rm KH}$ exists~\cite{BouHaw96}. In our case, by contrast, the normalization is set by the choice of evolution parameter $R$, which is ambiguous. 

In Sec.~\ref{sec-examples} we will consider a particularly simple choice of parametrization. Remarkably, we will find for a large class of dynamical solutions that the acceleration defined in Eq.~(\ref{k_off_H}) agrees with the standard Killing surface gravity of the corresponding static solutions.

\subsection{Gauge and Reparametrization Transformations}
\label{sec-symm}

There are two kinds of transformations that do not change the screen and preserve the conventions of Eqs.~(\ref{eq-kl}) and (\ref{eq-hR}). The first transformation is analogous to a global symmetry, in that it does not depend on the position. The second is a $U(1)$ gauge symmetry.

The first symmetry is a trivial reparametrization of the label $R$ of the leaves. There are certain geometrically motivated choices one could consider in order to fix $R$: for example, by linking it to the area $A$ of the leaves, e.g.\ via $A=4\pi R^2$ or $A=\exp(R)$. Here we will insist only that $R$ grow monotonically with $A$. Then we can consider any transformation $R \to R'$ with
\begin{equation}
\exp[\gamma(R)] \equiv \frac{dR'}{dR}>0~.
\end{equation}
Note that $\gamma$ can only depend on $R$, not on the angular position on each leaf. The above conventions and definitions imply the following transformation properties: %(CORRECTED THE TRANSFORMATION LAW OF $\tilde{\kappa}$--PLEASE CHECK.).
\begin{eqnarray}
h & \to & e^{-\gamma} h\\
n & \to & e^{-\gamma} n\\
l & \to & e^{-\gamma} l\\
k & \to & e^\gamma k\\
\beta & \to & e^{-2\gamma} \beta \\
\Omega_a & \to & \Omega_a\\
\tilde\kappa & \to & e^{-\gamma} (\tilde\kappa + \gamma'(R) )~.
\end{eqnarray}
The extrinsic curvature tensors, $B_{ab}^{(h,n,k,l)}$, and their components (expansion and shear), transform like $h,k,n,l$, respectively.

A second symmetry arises from rescaling $\alpha$ by an arbitrary positive function of $R$ and of the angular position, while holding $h$, $n$, and $R$ fixed. This requires taking 
$l \to e^{-\Gamma} l$, and by Eq.~(\ref{eq-kl}), $k\to e^\Gamma k$.
The remaining screen parameters transform as\footnote{Note that $(\tilde\kappa, \Omega_a)$ transform like $(A_0,\mathbf{A})$, the electric and magnetic potential, under a gauge transformation $\Gamma$. It would be nice to relate this to a shift by $\Gamma$ in the phase of a nonrelativistic wavefunction $\psi$~\cite{Son-13,Jensen-14,Geracie-15} that is part of the quantum gravity theory on the screen.}
\begin{eqnarray}
\alpha & \to & e^\Gamma\alpha\\
\beta & \to & e^{-\Gamma} \beta\\
\tilde\kappa & \to & \tilde\kappa+\dot\Gamma\\
\Omega_a & \to & \Omega_a+D_a\Gamma~.
\end{eqnarray}
Note that the combination
\begin{eqnarray} 
\widehat{\Omega}_{a} & \equiv &  h^{b}q_{a}^{\mbox{ } c} \nabla_{c}n_{b} \, , \label{eq-def-tilde-omega}\\
& = &  -2\alpha\beta \Omega_{a} + \beta D_{a}\alpha - \alpha D_{a}\beta \, .  \label{eq-exp-tilde-omega}
\end{eqnarray} 
is invariant under the gauge symmetry.

Again, it is possible to gauge-fix this symmetry. For example, we can insist that $\alpha=-1$ everywhere, or that $\theta^{(l)}=-1$. Below we find that different choices are convenient in different applications. However, the most general evolution equations we display in the next section will be invariant under any of the above transformations.

\section{Dynamics and Observer Dependence}
\label{sec-dod}

A holographic screen is a codimension-one hypersurface in spacetime. Hence, it must obey the constraint equations of General Relativity, 
\begin{equation}
G_{ab} n^b = 8\pi  T_{ab} n^b~.
\label{eq-gr}
\end{equation}
These four equations are usually expanded in a $3+1$ formalism, as one energy constraint plus three momentum constraints on the 3-metric and 3-extrinsic curvature. 

Here we are dealing with a hypersurface of indefinite signature, but with the additional structure of a 2+1 decomposition, the foliation into leaves. Thus it is natural to express Eq.~(\ref{eq-gr}) in terms of the kinematic quantities defined in the previous section, which are adapted to this foliation. One finds %XXX Q INSTEAD
\begin{widetext}
\begin{eqnarray}
\alpha (\widehat{\mathcal{L}}_{h}+\tilde\kappa)\theta^{(l)} + D_{a}\widehat{\Omega}^{a} & = & 8\pi T_{ab}n^{a}h^{b} + B^{(h)}_{ab}B_{(n)}^{ba} \label{eq-kc} \\
(\widehat{\mathcal{L}}_{h}+\theta^{(h)})\Omega_{c}- D_{c}\tilde\kappa + \alpha D_c\theta^{(l)}
 & = &  8\pi T_{ab}n^{a}q^{b}_{c} - D_{a}B_{c}^{(n)a} \label{eq-omegadyn} \\
-\frac{\alpha}{2} \mathcal{R} +\alpha \Omega_a\Omega^a  -\alpha D_{a}\Omega^{a} -2\Omega^aD_a\alpha+D_aD^a\alpha  & = & 8\pi T_{ab} n^a k^b  +\beta\sigma^{(k)}_{ab}\sigma_{(k)}^{ab}~,
\label{eq-bc}
\end{eqnarray} 
%\end{widetext}
where we have used $\theta^{(k)}=0$ to simplify the equations. Recall that $\widehat{\Omega}_{a}$ is not an independent variable but given by Eq.~(\ref{eq-exp-tilde-omega}). Here $\mathcal{R}$ is the Ricci scalar associated with the leaf metric, $q_{ab}$. In addition, there is a dynamical equation for the metric on the leaves,
\begin{equation}
\widehat{\mathcal{L}}_{h} q_{ab} = B^{(h)}_{ab}~.
\label{eq-qb}
\end{equation}
The evolution operator $\widehat{\mathcal{L}}_{h}$ acts on the tensors which are purely tangent to the leaf as~\cite{Gourgoulhon-TH}
\begin{equation}
\widehat{\mathcal{L}}_{h} A_{ab\ldots c} \equiv ~ q_{a}^{\, a'}q_{b}^{\, b'}\ldots q_{c}^{\, c'}  \, \mathcal{L}_{h} A_{a'b'\ldots c'} \, ,
\end{equation}
where we consider $\mathcal{L}_{h}$ as an operator on $H$.

This system of equations is invariant under the symmetries described in Sec.~\ref{sec-symm}. We have displayed intrinsic quantities associated with the screen on the left side. Extrinsic quantities that act like sources appear on the right hand side. We will now describe three ways in which one might interpret this system of equations, using different gauge choices.

\subsection{Viscous Fluid Analogy}
\label{sec-dod1}

We begin by regarding both the spacetime $M$ and the screen $H$ as fixed. In this case we are merely expressing the 3D intrinsic and extrinsic curvatures of $H$ as the evolution of 2D screen variables along $H$. This may nevertheless be interesting if it throws new light on the system. In fact, the evolution equations bear some similarity to fluid equations. We will identify a number of problems with the fluid interpretation, however.

To obtain fluid-like equations, we will set $\alpha=-1$ to gauge-fix the $U(1)$ symmetry. We do not gauge-fix the screen parameter $R$. Equations~(\ref{eq-kc}--\ref{eq-qb}) become
\begin{eqnarray}
(\widehat{\mathcal{L}}_{h}+\theta^{(h)})\theta^{(h)}  +(\tilde\kappa-\theta^{(h)})\theta^{(h)} - B^{(h)}_{ab}B_{(n)}^{ba} + D_{a}\widehat{\Omega}^{a} & = & 8\pi T_{ab}n^{a}h^{b} \label{dynamical_eq_theta-com} \\
(\widehat{\mathcal{L}}_{h}+\theta^{(h)})\Omega_{c}- D_{c}(\tilde\kappa-\theta^{(h)}) +D_{a}B_{c}^{(n)a}
 & = &  8\pi T_{ab}n^{a}q^{b}_{c} \label{NS_omega-com} \\
\frac{1}{2} \mathcal{R} - \Omega_a\Omega^a  + D_{a}\Omega^{a}  -\beta\sigma^{(k)}_{ab}\sigma_{(k)}^{ab}  & = & 8\pi T_{ab} n^a k^b~,\label{eq-bc2}\\
\widehat{\mathcal{L}}_{h} q_{ab} & = & B^{(h)}_{ab}~.\label{eq-qb2}
\end{eqnarray} 
%\end{widetext}
We expand the extrinsic curvature terms using Eqs.~(\ref{eq-expansion}) and (\ref{eq-shear}), to obtain\footnote{The equations appear slightly simpler than in Ref.~\cite{Gourgoulhon-TH,Gourgoulhon-DH,Gourgoulhon-DH-1} due to a difference in conventions. There, the evolution vector $h$ satisfies $h_a l^a=1$ (in our notation). This convention is not well-defined when $\beta=0$, i.e., at points where the screen changes signature. The convention we adopt in this subsection, $h_a k^a=1$, is everywhere well-defined; this follows from the area theorem~\cite{BouEng15a}.}
\begin{eqnarray} 
\widehat{\mathcal{L}}_{h}\theta^{(h)} + \theta^{(h)2} & = & -\tilde\kappa\theta^{(h)} + \frac{1}{2}\theta^{(h)2} + \sigma^{(h)}_{ab}\sigma^{(n)ba} - D_{a}\widehat{\Omega}^{a} + 8\pi T_{ab}h^{a}n^{b} \label{dynamical_eq_theta}\\
\widehat{\mathcal{L}}_{h}\Omega_{c} + \theta^{(h)}\Omega_{c} & = & D_{c}(\tilde\kappa) -D_{a}\sigma_{c}^{(n)a} - \frac{1}{2}D_{c}\theta^{(h)} + 8\pi T_{ab}n^{a}\tilde{q}^{b}_{c} \label{NS_omega}\\
-\frac{1}{2} \mathcal{R} + \Omega_a\Omega^a  -D_{a}\Omega^{a}  & = & 8\pi T_{ab} n^a k^b  +\beta\sigma^{(k)}_{ab}\sigma_{(k)}^{ab}~,\label{eq-bc3}\\
\widehat{\mathcal{L}}_{h} q_{ab} & = & \frac{1}{2} \theta^{(h)} q_{ab} + \sigma^{(h)}_{ab}~.\label{eq-qb3}
\end{eqnarray} 
\end{widetext}

With the definitions
\begin{align}
	\Pi_{c} \equiv\mbox{ }&-\frac{1}{8 \pi}\Omega_{c}\mbox{ }\mbox{ }\mbox{ }\mbox{ }\mbox{ }\mbox{ }\mbox{ }\mbox{ }\mbox{ }\mbox{ }\mbox{ }\mbox{ }\mbox{ }\mbox{ }\mbox{ }\text{momentum density}\\
	\epsilon \equiv\mbox{ }&\frac{1}{8\pi}\theta^{(h)}\mbox{ }\mbox{ }\mbox{ }\mbox{ }\mbox{ }\mbox{ }\mbox{ }\mbox{ }\mbox{ }\mbox{ }\mbox{ }\mbox{ }\mbox{ }\mbox{ }\mbox{ }\mbox{ }\mbox{ }\text{energy density}\\
	P \equiv\mbox{ }& \frac{1}{8 \pi} (\tilde\kappa-\theta^{(h)})\mbox{ }\mbox{ } \, \, \mbox{ }\mbox{ }\mbox{ }\mbox{ }\mbox{ }\mbox{ }\text{pressure}\\
		Q_{c} \equiv\mbox{ }&\frac{1}{8\pi}\widehat{\Omega}_{c}\mbox{ }\mbox{ } \, \, \, \mbox{ }\mbox{ }\mbox{ }\mbox{ }\mbox{ }\mbox{ }\mbox{ }\mbox{ }\mbox{ }\mbox{ }\mbox{ }\mbox{ }\mbox{ }\mbox{ }\mbox{ }\text{heat current}
\end{align}
\mbox{}
\begin{align}
	\zeta \equiv\mbox{ }& \frac{1}{16\pi}\mbox{ }\mbox{ }\mbox{ }\mbox{ }\mbox{ }\mbox{ }\mbox{ } \, \mbox{ }\mbox{ }\mbox{ }\mbox{ }\mbox{ }\mbox{ }\mbox{}\mbox{ }\mbox{ }\mbox{ }\mbox{ }\mbox{ }\mbox{ }\text{bulk viscosity}\\
\mu \equiv\mbox{ }& \frac{1}{8\pi}\mbox{ }\mbox{ }\mbox{ }\mbox{ }\mbox{ }\mbox{ }\mbox{ }\mbox{ }\mbox{ }\mbox{ }\mbox{ }\mbox{ }\mbox{ }\mbox{ }\mbox{}\mbox{ }\mbox{ }\mbox{ }\mbox{ }\mbox{ }\mbox{ }\mbox{ }\text{shear viscosity}\\
	f_{c} \equiv\mbox{ }&-T_{ab}n^{a}\tilde{q}^{b}_{c}\mbox{ }\mbox{ }\mbox{ }\mbox{ }\mbox{ }\mbox{ }\mbox{ }\mbox{ }\mbox{ }\mbox{ }\mbox{ }\text{external force density}\\
q \equiv\mbox{ }&T_{ab}n^{a}h^{b}\mbox{ }\mbox{ }\mbox{ }\mbox{ }\mbox{ }\mbox{ } \, \mbox{ }\mbox{ }\mbox{ }\mbox{ }\mbox{ }\mbox{ }\mbox{ }\mbox{ }\text{external heat source}
\end{align}
\begin{widetext}
\noindent Eq.~\eqref{NS_omega} resembles the Navier-Stokes equation for the momentum density
\begin{equation}
\widehat{\mathcal{L}}_{h}\Pi_{c} + \theta^{(h)}\Pi_{c} = -\mbox{ }D_{c}P + \mu D_{a}\sigma^{(n)a}_{c} + \zeta D_{c}\theta^{(n)} + f_{c}~;
\label{eq-navstok}
\end{equation}
and Eq.~\eqref{dynamical_eq_theta} resembles an equation governing the flux of the internal energy:
\begin{equation} 
	\widehat{\mathcal{L}}_{h}\epsilon + \theta^{(h)}\epsilon = -P\theta^{(h)} + \zeta\theta^{(h)}\theta^{(n)}  + \mu\sigma^{(h)}_{ab}\sigma^{(n)ba} -D_{a}Q^{a} + q \label{eq-enflux}
\end{equation} 
\end{widetext}

First, let us note that the bulk viscosity is positive. This contrasts with the negative (hence unstable) bulk viscosity of the event horizon fluid in the membrane paradigm of Price and Thorne~\cite{price-thorne}. This is simply because we absorbed an addition term proportional to $\theta^{(h)}$ into the definition of the pressure. With an analogous definition of pressure, one would also find a positive bulk viscosity in~\cite{price-thorne}. We do not regard this as a success, however. Rather, the fact that pressure and bulk viscosity terms cannot be uniquely identified is a first sign that the fluid analogy fails. We will discuss additional problems below.

Note that Refs.~\cite{Gourgoulhon-TH,Gourgoulhon-DH,Gourgoulhon-DH-1} also obtain a positive bulk velocity, but for a different reason: by defining the pressure to be $\tilde\kappa$, and taking $\theta^{(h)}$ rather than $\theta^{(n)}$ to be the expansion rate relevant to the bulk viscosity. However, the same tensor should define both the expansion and the shear. Since $\sigma^{(n)}$ appears in the shear viscosity term of Eq.~(\ref{eq-navstok}), we require that $\theta^{(n)}$, and not $\theta^{(h)}$, appear in the bulk viscosity term. Yet, this requirement, too, appears inconsistent, since the expansion that controls the dilution of the energy and momentum densities is $\theta^{(h)}$. 

These ambiguities and contradictions lead us to recognize that the viscous fluid analogy has multiple, serious shortcomings:
\begin{itemize} 
\item There is no equation of state that would determine the pressure $\tilde\kappa$ from other dynamical parameters intrinsic to the fluid.
\item There is no dynamical equation for the number density or mass density of fluid particles, analogous to the continuity equation.
\item Therefore, there is no well-defined velocity vector field (``$v^b$''). 
\item Therefore, the rate of shear and expansion cannot be computed from the dynamical equations (via ``$D_a v^b$''). Rather, these rates are an arbitrary external input variable.
\item The dissipation term $\mu\sigma^{(h)}_{ab}\sigma^{(n)ba}$ in Eq.~(\ref{eq-enflux}) corresponds neither to a Newtonian, nor properly to a non-Newtonian fluid. $\sigma^{(h)}_{ab}$ is entirely independent of $\sigma^{(n)}_{ab}$, so the viscous stresses are not a function of fluid variables alone. 
\end{itemize} 
Some of this criticism also applies to the fluid description of event horizons in the membrane paradigm~\cite{Damour,price-thorne}, as has also been noted by Strominger and collaborators~\cite{BreKee11}.

Finally, it is not clear what the interpretation of the remaining Eqs.~(\ref{eq-bc3}) and (\ref{eq-qb3}) is, in the fluid picture. They state that not all external input parameters are completely independent, such as $\sigma_{ab}^{(k)}$ and $\sigma_{ab}^{(h)}$, $q_{ab}$ and $T_{nk}$. Alternatively we may regard Eq.~(\ref{eq-bc3}) as a constraint equation determining the parameter $\beta$.
%, which appears in the (nonstandard) relation between the momentum density, $\Omega^a$, and the heat current or energy flux density, $Q^a$. 

To conclude, we do not find the interpretation of screen evolution as fluid dynamics to be plausible. Moreover, the above analysis, with $M$ and $H$ fixed, actually ignores a crucial degree of freedom, as we shall see next.

\subsection{Observer-Dependence}
\label{sec-dod2}

An instructive way to think about the evolution equations is to consider only the 4D spacetime $M$ as given. Our task is to construct a holographic screen, $H$. Once we have started, the equations tell us how to find the (infinitesimally) next leaf. 

This task is ambiguous, because each leaf is associated with a null slice, there are many ways of picking a null foliation of $M$. We can regard $\alpha<0$ as a free parameter that determines a choice of a null foliation (for a fixed, arbitrary choice of null vector field $l$ at every leaf). There is no equation determining $\alpha$, because it is a genuine ambiguity, corresponding to the ``observer-dependence'' of holographic screens. 

Let us define an effective stress tensor %(ADDED A FACTOR OF $8\pi$)
\begin{widetext}
\begin{eqnarray} 
8\pi \bar T_{ab} & \equiv & 8\pi T_{ab} + k_a k_b B_{(l)}^{cd} B^{(l)}_{cd}+ l_a l_b B_{(k)}^{cd} B^{(k)}_{cd}\\
& = & 8\pi T_{ab} + k_a k_b \left(\frac{\theta_{(l)}^2}{2}+\sigma_{(l)}^{cd} \sigma^{(l)}_{cd}\right)
+ l_a l_b \sigma_{(k)}^{cd} \sigma^{(k)}_{cd}~.
\label{eq-tbardef}
\end{eqnarray} 
This takes a form similar to the effective stress-energy of gravitational radiation in linearized gravity. In general no local definition of energy can be given for gravitational degrees of freedom, but here the holographic screen provides additional structure analogous to a preferred background. Thus, $\bar T_{ab}$ can be interpreted as incorporating stress energy associated with gravitational radiation crossing the leaf orthogonally.\footnote{In~\cite{Hayward-grav-tensor}, Hayward derives a stress tensor for the gravitational radiation in a ``quasi-spherical'' approximation. We do not work in this approximation, but we not that his result takes the same form as our definition in Eq.~(\ref{eq-tbardef}).}

Thus Eqs.~(\ref{eq-kc}--\ref{eq-bc}) become 
\begin{eqnarray}
\alpha (\widehat{\mathcal{L}}_{h}+\tilde\kappa)\theta^{(l)} + D_{a}(-2\alpha\beta\Omega^{a} + \beta D^{a}\alpha -\alpha D^{a}\beta)  & = & 8\pi \bar T_{ab}n^{a}h^{b} \label{eq-kced} \\
(\widehat{\mathcal{L}}_{h}+\theta^{(h)})\Omega_{c}- D_{c}\tilde\kappa + \alpha D_c\theta^{(l)}
 & = &  8\pi \bar T_{ab}n^{a}q^{b}_{c} - D_{a}B_{c}^{(n)a} \label{eq-omegadyned} \\
-\frac{\alpha}{2} \mathcal{R} +\alpha \Omega_a\Omega^a  -\alpha D_{a}\Omega^{a} -2\Omega^aD_a\alpha+D_aD^a\alpha + 8\pi\alpha \bar T_{ab} k^a l^b & = & 8\pi \beta \bar T_{ab} k^a k^b~,
\label{eq-bced}
\end{eqnarray} 
\end{widetext}
Eq.~(\ref{eq-qb}) is trivial from this viewpoint, so we have not listed it again.

Geometrically, we can think of the role of $\alpha$ and $\beta$ by considering the forward evolution of the screen by an infinitesimal ``time'' step $dR$ (see Fig.~\ref{figure-ab}). In order to find the next leaf after $\sigma(R)$, we transport the leaf $\sigma(R)$ infinitesimally along $\alpha l$ to a nearby surface $\bar\sigma(R+dR)$.  In general this surface will not be marginally trapped, but it does define a new null slice, $N(R+dR)$, generated by the $k$-lightrays orthogonal to $\bar\sigma(R+dR)$. Then we find the cut with $ \theta^{(k)}=0$ on $N(R+dR)$. This gives the new leaf $\sigma(R+dR)$.

Eq.~(\ref{eq-bced}) can be regarded as a constraint equation that allows us to short-circuit this construction. It can be solved for $\beta$, because the generic condition of Refs.~\cite{BouEng15a,BouEng15b} requires that \begin{equation}
\bar T_{ab} k^a k^b > 0~. \label{eq-generic}
\end{equation}
Then $\sigma(R+dR)$ can be obtained directly, by transporting the surface $\sigma(R)$ along the vector $h=\alpha l +\beta k$. Thus, the parameter $\beta$ tells us how far to slide up or down $N(R+dR)$ to get from the ``wrong'' surface $\bar\sigma(R+dR)$ to get to the correct new leaf $\sigma(R+dR)$. 

The remaining Eqs.~(\ref{eq-kced}) and (\ref{eq-omegadyned}) describe the evolution of the vector fields $k$ and $l$ which are linked by the condition $k^{a}l_{a}=-1$. They provide additional structure beyond the given spacetime $M$, associated with the screen $H$. As shown in Appendix~\ref{sec-partran}, the failure of $k$ and $l$ to be parallel-transported into themselves along $H$ by $h$ is captured by $\tilde\kappa$, $\alpha$, $\beta$, and the vector field $\Omega_c$: 
\begin{eqnarray} 
h^{b}\nabla_{b}k_{a} &  = &  \tilde\kappa k_{a} + D_{a}\alpha - \alpha\Omega_{a}  \label{eq-kder}\\
h^{b}\nabla_{b}l_{a}  & = &  -\tilde\kappa l_{a} + D_{a}\beta + \beta \Omega_{a}\label{eq-lder}
\end{eqnarray} 
Note that both $\theta^{(l)}$ and its derivative are fully determined by the arbitrary choice of the ``length'' of $l$ at each leaf. Here we take the ``length'' of $l$ as input, so Eq.~(\ref{eq-kced}) acts as a constraint that determines $\tilde\kappa$. [Alternatively, we could specify $\tilde{\kappa}$ and thus fix the length of $l$ via Eq.~(\ref{eq-kced}).] Finally, Eq.~(\ref{eq-omegadyned}) is a dynamical evolution equation for $\Omega_c$.

\subsection{Background-Free Description}
\label{sec-dod3}

Finally, we consider an interpretation where neither $M$ nor $H$ are given. Then we may regard Eqs.~(\ref{eq-kc}--\ref{eq-qb}) as a nonrelativistic system evolving with the time variable $R$. The advantage of this viewpoint is that it makes no reference to the spacetime that the screen is embedded in, or even to an induced 2+1 metric on the screen. This minimal approach may be appropriate if we regard the screen as a (partially) pre-geometric object that arises from an underlying quantum gravity theory in an appropriate regime. It may be natural for the screen to be constructed as a first step, before reconstructing the entire 4D geometry and fields. Eqs.~(\ref{eq-kc}--\ref{eq-qb}) constrain this construction.

In this case it is convenient to choose a gauge in which $\theta^{(l)}=-1$, so that Eqs.~(\ref{eq-kc}--\ref{eq-qb}) reduce to
\begin{widetext}
\begin{eqnarray}
-\alpha \tilde\kappa -  
D_a\left[ \alpha\beta \left(2\Omega^a+D^a\log\frac{\beta}{\alpha} \right) \right] & = & 8\pi \bar T_{ab}n^a h^b\\
%\alpha\left(\dot\Omega_c+\Omega_c\right) - D_c\tilde\kappa & = & T_{ab}n^{a}q^{b}_{c}  - D_a\sigma_c^{(n)a}\\
\dot\Omega_c -\alpha\Omega_c - D_c\tilde\kappa + \frac{1}{2} D_{c}\alpha & = & 8\pi \bar T_{ab}n^{a}q^{b}_{c}  - D_a\sigma_c^{(n)a}\\
\alpha\left[\Omega_a\Omega^a-D_a\Omega^a-2\Omega^aD_a \log\alpha - \frac{\cal R}{2}+
\alpha^{-1}D_aD^a\alpha \right] & = & 8\pi \bar T_{ab}n^a k^b \\ 
\dot q_{ab} & = & -\frac{\alpha}{2} q_{ab} + \alpha \sigma^{(l)}_{ab} + \beta \sigma^{(k)}_{ab}~.
\end{eqnarray}
\end{widetext}
We have replaced the Lie-derivatives with dots, since in this viewpoint they are simple time derivatives. Objects such as $k,l,h,n$ are now considered to emerge in the reconstruction of the geometry. For example, the length of integral curves of $h$ is related to the evolution parameter $R$ by $(dL/dR)^2 = -2\alpha\beta$, where positive values correspond to a spacelike signature. Similarly, $\tilde\kappa$ and $\Omega$ allow us to reconstruct the null vector fields $k$ and $l$ by integration of Eqs.~(\ref{eq-kder}) and (\ref{eq-lder}). None of these geometric concepts are intrinsic to the above equations, but they can be reconstructed from them. 

We may regard $\alpha,\beta,\kappa,\Omega_c$, and the 2D metric $q_{ab}$ as intrinsic quantities of the holographic screen, but they are highly underdetermined. It is not clear whether $\sigma_{ab}^{(k,l)}$ and $T_{ab}n^b$ are best regarded as input (which happens to correspond to the matter stress tensor and gravitational waves in the reconstructed 4D spacetime); or rather whether the above equations should be viewed as determining certain components of the stress tensor and the shear, given arbitrary input for the screen quantities $\alpha,\beta,\kappa,\Omega_c$. One parameter (most naturally $\alpha$) is associated with a null foliation of the 4D spacetime. For each leaf of the screen, microscopic data should determine the quantum state on the associated null slice. 

\section{Examples of Holographic Screens}
\label{sec-examples}

In this section, we work out a number of detailed examples of physical interest. Several of the holographic screens we will construct are spherically symmetric. Therefore, we will begin by listing general results that apply to all spherical screens, before specializing further.

\subsection{Implications of Spherical Symmetry}
\label{sec-ss}

Consider a screen $H$ embedded in a spacetime $M$, such that both are invariant under spherical symmetry. In this case we shall choose the area radius as the evolution parameter $R$,
\begin{equation}
A=4\pi R^2~.
\end{equation}
We shall further choose the convention that 
\begin{equation}
\alpha=-1~,
\end{equation}
which can be regarded as gauge-fixing the rescaling symmetry of $l$. The metric $q_{ab}$ is of the form
\begin{equation}
q_{ab} = R^2 s_{ab}
\end{equation}
where $s_{ab}$ is the metric on the unit two-sphere. Using the above conventions of $R$ and $\alpha$, one finds
\begin{equation}
\theta^{(l)} = \theta^{(n)} = - \theta^{(h)}  = - \frac{d}{dR}\log A = -\frac{2}{R}~.
\end{equation}
The shears and the normal one form would break spherical symmetry and so must vanish, 
\begin{equation}
\sigma_{cd}^{(k,l,h,n)} = 0~,~~\Omega_c =0 ~.
\end{equation}
Since $h^a h_a = 2\beta$, the induced 3-metric on $H$ is
\begin{equation}
ds_{H}^{2} =  2\beta dR^{2} + R^{2}d\Omega^{2}~;
\end{equation}
Again, this is only well-defined piecewise on portions with definite sign of $\beta$, and we will not consider this metric further.

\begin{figure*}[t]
\begin{tikzpicture}[scale=0.45]
\node at (4.5,10) {$w=0$};
\node at (14.5,10) {$w=1/3$};
\node at (24.5,10) {$w= - 9/10$};
	\draw [dashed](0,0) --(9,0);
	\draw [black,thick](0,0) --(0,9);
	\draw [black](0,9) --(9,0);
	\draw [dotted](0,8) --(8,0);
	\draw [dotted](0,7) --(7,0);
	\draw [dotted](0,6) --(6,0);
	\draw [dotted](0,5) --(5,0);
	\draw [dotted](0,4) --(4,0);
	\draw [dotted](0,3) --(3,0);
	\draw [dotted](0,2) --(2,0);
	\draw [dotted](0,1) --(1,0);
	\fill [black](0.89,7.11) circle (.1cm);
	\fill [black](1.38,5.62) circle (.1cm);
	\fill [black](1.5,4.5) circle (.1cm);
	\fill [black](1.42,3.58) circle (.1cm);
	\fill [black](1.21,2.78) circle (.1cm);
	\fill [black](0.96,2.04) circle (.1cm);
	\fill [black](0.65,1.35) circle (.1cm);
	\fill [black](0.33,0.67) circle (.1cm);
	\draw [blue](0,0) --(0.33,0.67) --(0.65,1.35) --(0.96,2.04) --(1.21,2.78) --(1.42,3.58) --(1.5,4.5) --(1.38,5.62) --(0.89,7.11) --(0,9);
	\draw [blue](1.06,3.56) to (1.46,3.96) to (1.86,3.56);
	%	\node at (-.7,4.5) {$r=0$};
		\node at (4.5,-.5) {$t=0$};
		\node at (4,5.9) {$\mathcal{I}^{+}$};
		\draw [dashed](10,0) --(19,0);
		\draw [black,thick](10,0) --(10,9);
		\draw [black](10,9) --(19,0);
		\draw [dotted](10,8) --(18,0);
	\draw [dotted](10,7) --(17,0);
	\draw [dotted](10,6) --(16,0);
	\draw [dotted](10,5) --(15,0);
	\draw [dotted](10,4) --(14,0);
	\draw [dotted](10,3) --(13,0);
	\draw [dotted](10,2) --(12,0);
	\draw [dotted](10,1) --(11,0);
	\fill [black](10.5,0.5) circle (.1cm);
	\fill [black](11,1) circle (.1cm);
	\fill [black](11.5,1.5) circle (.1cm);
	\fill [black](12,2) circle (.1cm);
	\fill [black](12.5,2.5) circle (.1cm);
	\fill [black](13,3) circle (.1cm);
	\fill [black](13.5,3.5) circle (.1cm);
	\fill [black](14,4) circle (.1cm);
	\draw [blue](10,0) --(14.5,4.5);
	\draw [blue](12.25,2.75) to (12.75,2.75) to (12.75,2.25);
	\node at (14.5,-.5) {$t=0$};
		\node at (14,5.9) {$\mathcal{I}^{+}$};
		%%%%%%%%%%%%%%%%%%%%%%%%%%%%%%%%%%%%%%%%%%
			\draw [dashed](20,0) --(29,9);
	\draw [black,thick](20,9) --(20,0);
	\draw [black](20,9) --(29,9);
	\draw [dotted](20,8) --(24,4);
	\draw [dotted](20,7) --(23.5,3.5);
	\draw [dotted](20,6) --(23,3);
	\draw [dotted](20,5) --(22.5,2.5);
	\draw [dotted](20,4) --(22,2);
	\draw [dotted](20,3) --(21.5,1.5);
	\draw [dotted](20,2) --(21,1);
	\draw [dotted](20,1) --(20.5,0.5);
	\draw [red](20,9) --(24.5,4.5);
	\fill [black](22.77,5.23) circle (.1cm);
	\fill [black](22.87,4.13) circle (.1cm);
	\fill [black](22.60,3.40) circle (.1cm);
	\fill [black](22.22,2.78) circle (.1cm);
	\fill [black](21.81,2.19) circle (.1cm);
	\fill [black](21.37,1.63) circle (.1cm);
	\fill [black](20.92,1.08) circle (.1cm);
	\fill [black](20.46,0.54) circle (.1cm);
	\draw [blue](20,9) --(20.56,8.34) --(21.07,7.73) --(21.49,7.21) --(21.84,6.76) --(22.11,6.39) --(22.32,6.08) --(22.48,5.82) --(22.60,5.60) --(22.69,5.40) --(22.77,5.23) --(22.91,4.59) --(22.87,4.13) --(22.60,3.40) --(22.22,2.78) --(21.81,2.19) --(21.37,1.63) --(20.92,1.08) --(20.46,0.54) --(20,0);%--(21.93,2.07) --(21.46,1.54) --(20.97,1.03) --(20.49,0.51) --(20,0) ;
	\draw [blue](22.47,4.28) to (22.87,4.68) to (23.27,4.28);
	%\draw [green](0,0) --(0.10,-0.20) --(0.20,-0.40) --(0.30,-0.60) --(0.39,-0.79) --(0.49,-0.99) --(0.89,-1.89) --(1.38,-3.38) --(1.5,-4.5) --(1.42,-5.42)--(1.22,-6.22) --(0.96,-6.96) --(0.65,-7.65) --(0.33,-8.33) --(0,-9) ;
		%\node at (19.5,4.5) {$r=0$};
		\node at (26.5,5) {$t=0$};
		\node at (24.5,8.5) {$\mathcal{I}^{+}$};
 \end{tikzpicture}
\caption{Penrose diagrams for a spatially flat FRW universe dominated by matter (left) and radiation (middle). The right diagram is an approximation to de~Sitter spacetime; it contains a fluid with positive energy and equation of state close to that of vacuum energy. To construct a past holographic screen $H$, we consider the past light-cones (dotted lines) of a comoving observer at $r=0$ (left edge). The surfaces of maximum area on each of these light cones (black dots) are the leaves of the screen $H$ (blue curve). Note that $H$ approaches the event horizon (red line) at late times, in the near-de~Sitter case. We find that the surface gravity $\kappa$ approaches that of de Sitter space in the limit.}
\label{fig-FRW}
		\end{figure*}
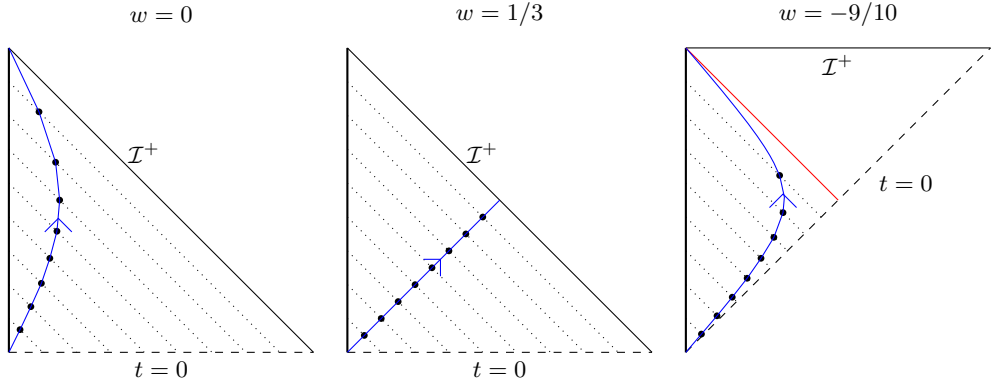
		
The only nontrivial intrinsic quantities associated with screen evolution are the slope, $\beta$, and the acceleration, $\kappa$. They are determined entirely by certain stress tensor components and by $R$, since Eqs.~(\ref{eq-kc}) and (\ref{eq-bc}) reduce to
\begin{eqnarray}
\tilde\kappa & = & 4\pi R ~ T_{ab}n^a h^b\\ \label{spherical-symm-kappa}
\beta & = & \frac{(8\pi R^2)^{-1}-T_{ab} k^a l^b}{T_{ab}k^a k^b}~. \label{spherical-symm-beta}
\end{eqnarray}
We have used ${\cal R} = 2/R^2$. The former equation is somewhat reminiscent of a first law, if we write it as
%The extrinsic curvature of $H$ in $M$ can be expressed in terms of other quantities we have defined; in the spherical case this takes a particularly simple form. We define XXX WHAT IF WE NORMALIZE N FIRST? OTHERWISE ADD A NOTE THAT THIS IS A NONSTANDARD DEFINITION?
% \begin{equation}
% \Theta_{ab} \cong\mbox{ }\gamma_{a}^{\mbox{ }c}\gamma_{b}^{\mbox{ }d}\nabla_{c}n_{d}, \label{ext_cur_screen_tab}
% \end{equation}
% which is diagonal in the case of spherical symmetric. That is, we write
% \begin{equation}
% \Theta_{ab} \cong\mbox{ }\frac{1}{2\beta}\theta_{\|}h_{a}h_{b} + \frac{1}{2}\theta_{\perp}q_{ab}, \label{ext_cur_screen_spherical_tab}
% \end{equation}
% where $\theta_{\|}$ is given as
% \begin{align}
% \theta_{\|} \cong\mbox{ }&\frac{1}{2\beta}h^{a}h^{b}\Theta_{ab}\\ 
% \cong\mbox{ }& \frac{1}{2\beta}h^{a}h^{b}\nabla_{b}n_{a}\\ 
% \cong\mbox{ }& \beta\kappa + \frac{1}{2}h^{a}\nabla_{a}\log\beta
% \end{align}
% where we have used \eqref{dn_h}. Similarly, we write $\theta_{\perp}$ as
% \begin{align}
% \theta_{\perp} \cong\mbox{ }&q^{ab}\Theta_{ab}\\
% \cong\mbox{ }& q^{ab}\nabla_{a}n_{b}\\
% \cong\mbox{ }& \theta^{(n)}\\
% \cong\mbox{ }& -\frac{2}{R}
% \end{align}
\begin{equation}
\frac{\tilde\kappa}{2\pi} \frac{d(A/4)}{dR} =\oint d^2\vartheta\sqrt{q}~T_{ab}n^{a}h^{b} ~. \label{conj}
\end{equation}
The equation for $\beta$ can also be written as a constraint linking the radius to a stress tensor component:
\begin{equation}
\frac{1}{8\pi R^{2}} = T_{ab}n^{a}k^{b}~.
\end{equation} 

\subsection{Expanding Universe} 
\label{sec-frw}

Let $M$ be a flat Friedmann-Robertson-Walker universe with fixed equation of state $p=w \rho$, $-1<w\le 1$; see Fig.~\ref{fig-FRW}. The stress tensor is
\begin{equation}
T_{ab} = \rho t_{a}t_{b} + p (g_{ab}+t_{a}t_{b}) \, .
\end{equation}
The metric is
\begin{equation}
ds^{2} = -dt^{2} + a^{2}(t)\left(dr^{2} + r^2 d\Omega^{2}\right)
\end{equation}
with
\begin{equation}
a(t) =t^q
\end{equation}
and
\begin{equation}
q = \frac{2}{3}\frac{1}{1+w}~.
\end{equation}

To generate a null foliation of $M$, we consider the past light-cone of each point on the worldline $r=0$; see Fig.~\ref{fig-frw-nonsym}). On each cone, there is a cross-section of maximal area (since $A\to 0$ as the big bang is approached). This surface has vanishing expansion, $\theta^{(k)} =0$, by construction. The relevant spheres lie at $(r,t)$ subject to the condition~\cite{CEB2,RMP}
\begin{equation}
r\dot{a}(t) - 1 = 0. \label{screen}
\end{equation}
The proper area radius is $R = ra(t)$.  The future-directed outgoing congruence from any sphere in this geometry is obviously expanding, so this is a past holographic screen. Therefore~\cite{BouEng15a}, we have $\alpha>0$. 

In order for the screen to stay centered on the comoving worldline $r=0$, we must take $\alpha$ to be independent of angle, for example
\begin{equation}
\alpha = 1~.
%ARE~THERE~OTHER~DEVIATING~CONVENTIONS??
%h^{a}h_{a} & = & -2\beta \, ,
\end{equation} 
We will make a different, angle-dependent choice in Sec.~\ref{sec-nonsym} below, corresponding to the construction of a nonspherical screen in the same spacetime (see Sec.~\ref{sec-dod2}). 

The null normals $k,l$ satisfying $k^a l_a =-1$, $\theta^{(l)} = 2/R$ are
\begin{eqnarray} 
k^{a} & = & \left(\frac{\partial}{\partial t}\right)^{a} - \frac{1}{a(t)}\left(\frac{\partial}{\partial r}\right)^{a},\\
l^{a} & = & \frac{1}{2} \left(\frac{\partial}{\partial t}\right)^{a} + \frac{1}{2a(t)}\left(\frac{\partial}{\partial r}\right)^{a}.
\end{eqnarray} 
The vectors normal and tangent to the screen are $n=-l+\beta k$ and $h=l+\beta k$ with
\begin{equation}
\beta = q - \frac{1}{2} = \frac{1}{6}\, \frac{1-3w}{1+w}
\end{equation}
This implies, for example, that the screen is timelike in a matter-dominated universe ($q = 2/3$, $w = 0$, $\beta=1/6$) and null for a radiation-dominated universe ($q = 1/2$, $w = 1/3$, $\beta=0$). For stiffer fluids the screen will be spacelike.
\begin{figure}
\centering
\begin{tikzpicture}
	\draw [black](0,-4) --(0,1);
	\draw [black](0,-4) --(4,0);
	\draw [black](4,0) --(3,1);
	\draw [fill=gray!60!white](0,-4) to [out=70,in=270] (1.5,1) --(0,1) --(0,-4);
	\draw [fill=gray!10!white](0,-4) to [out=70,in=270] (1.5,1) --(3,1) --(4,0) --(0,-4);
	\draw [decorate,decoration=zigzag](0,1) --(3,1);
	\draw [red](3,1) --(0,-2);
	\draw [blue](0,1) to [out=280,in=150] (0.75,-0.45) to [out=330,in=200] (1.05,-0.45) to [out=20,in=225] (2.95,1);
	\draw [blue](.8,-.31) to (0.95,-0.46) to (0.8,-0.61);
	\node at (0,-4.2) {$i^{-}$};
	\node at (4.2,0) {$i^{0}$};
	\node at (-0.6,-1.5) {$r=0$};
	\draw [dotted](2.5,1) --(0,-1.5);
	\draw [dotted](2,1) --(0,-1);
	\draw [dotted](1.5,1) --(0,-.5);
	\draw [dotted](1,1) --(0,0);
	\draw [dotted](0.5,1) --(0,.5);
%	\draw [black](0,-9) --(0,5);
%	\draw [black](0,-9) --(9,0);
%	\draw [black](9,0) --(4,5);
%	\draw [fill=gray!60!white](0,-9) to [out=70,in=270] (2,5) --(0,5) --(0,-9);
%	\draw [fill=gray!20!white](0,-9) to [out=70,in=270] (2,5) --(4,5) --(9,0) --(0,-9);
%	\draw [decorate,decoration=zigzag](0,5) --(4,5);
%	\draw [dotted](4,5) --(0,1);
\end{tikzpicture}
\caption{Penrose diagram for collapsing dust. The dark-shaded region is the dense region, $r<r_{*}$. The light shaded region contains arbitrarily dilute matter to satisfy the generic condition. We construct a holographic screen $H$ using the future light-cones (dotted lines) of an observer at $r=0$. Note that $H$ changes signature and approaches the event horizon (red line) from the inside. We find that $\kappa$ approaches the Schwarzschild surface gravity there.}
\label{fig-OS}
\end{figure}
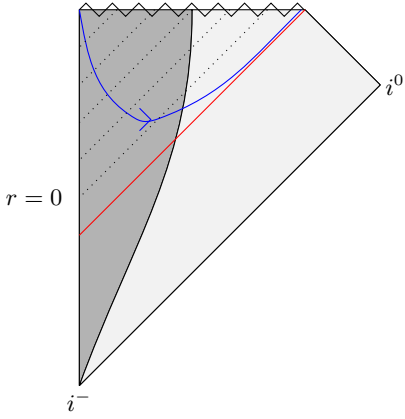

The screen acceleration is
\begin{equation}
\tilde\kappa = \frac{q-1}{R}~.
\end{equation}
The ``surface gravity'' defined in Eq.~(\ref{eq-kktb}) is
\begin{equation}
\kappa \equiv \frac{\tilde{\kappa}}{\beta} = \frac{2q-2}{2q-1}\frac{1}{R}~.
\end{equation}
For example, $\kappa = -2/R$ for the matter dominated universe. Notably, in the limit as $w \to -1$ ($q\to\infty$), this approaches the surface gravity of the de Sitter Killing horizon: $\kappa \to 1/R$.

\subsection{Collapsing Star}
\label{sec-os}
One can model a collapsing star by a finite, spherical, homogeneous dust ball. This is described by the Oppenheimer-Snyder solution~\cite{OppSny39}; see Fig.~\ref{fig-OS}. It can be constructed as a portion $r<r_*$ of a time-reversed Friedmann-Robertson-Walker cosmology, glued to a portion of the vacuum Schwarzschild solution. However, in order to satisfy the generic condition, Eq.~(\ref{eq-generic}), we will study the more general collapse of spherically symmetric dust with density $\rho(r)$. We can take $\rho(r)$ to become arbitrarily small outside some characteristic radius $r_*$. 

The holographic screens in such collapse scenarios were computed by Booth {\em et al.} in Ref.~\cite{Booth-examples}. Here we reproduce the relevant analysis and compute the screen quantities $\beta$ and $\tilde\kappa$.
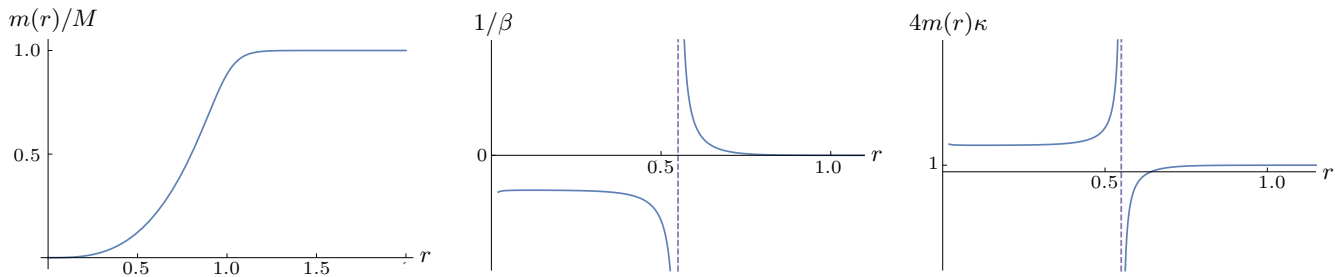
\begin{figure*}[t]
\begin{tikzpicture}
\node at (-8,0) {\includegraphics[scale=0.4]{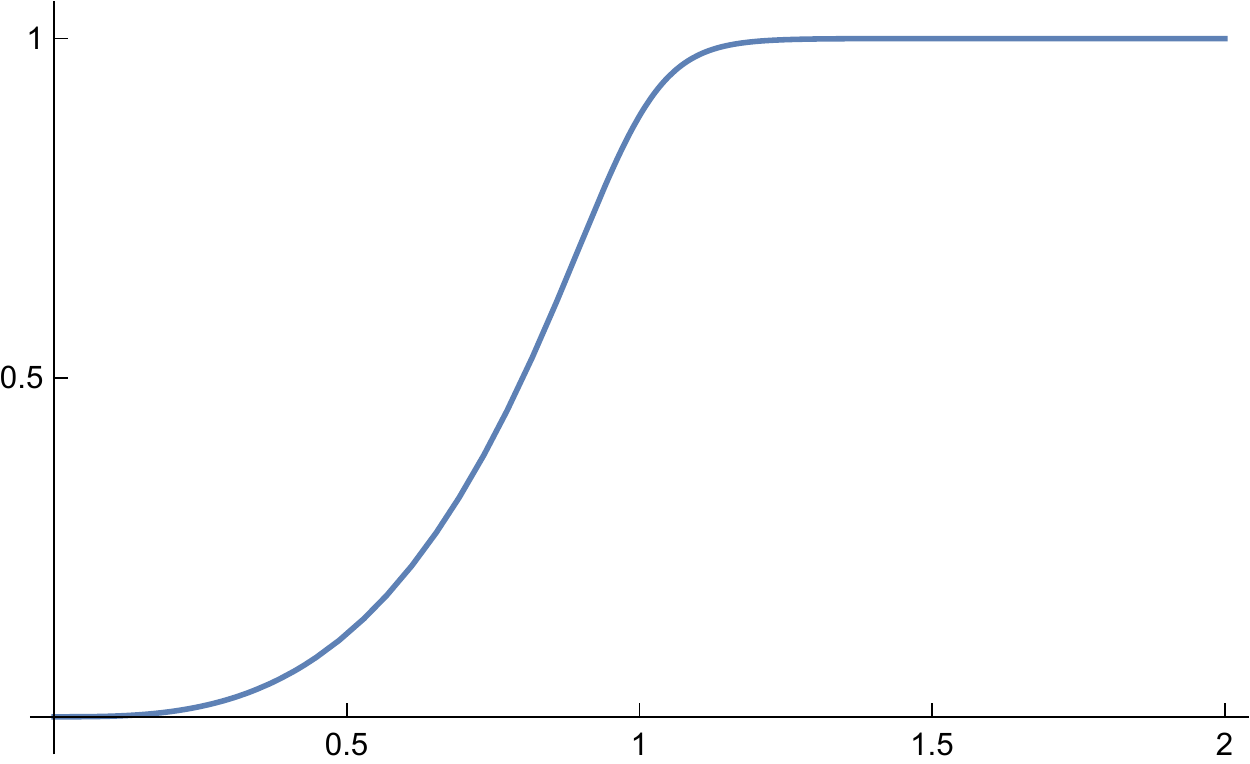}};
\node at (-2,0) {\includegraphics[scale=0.4]{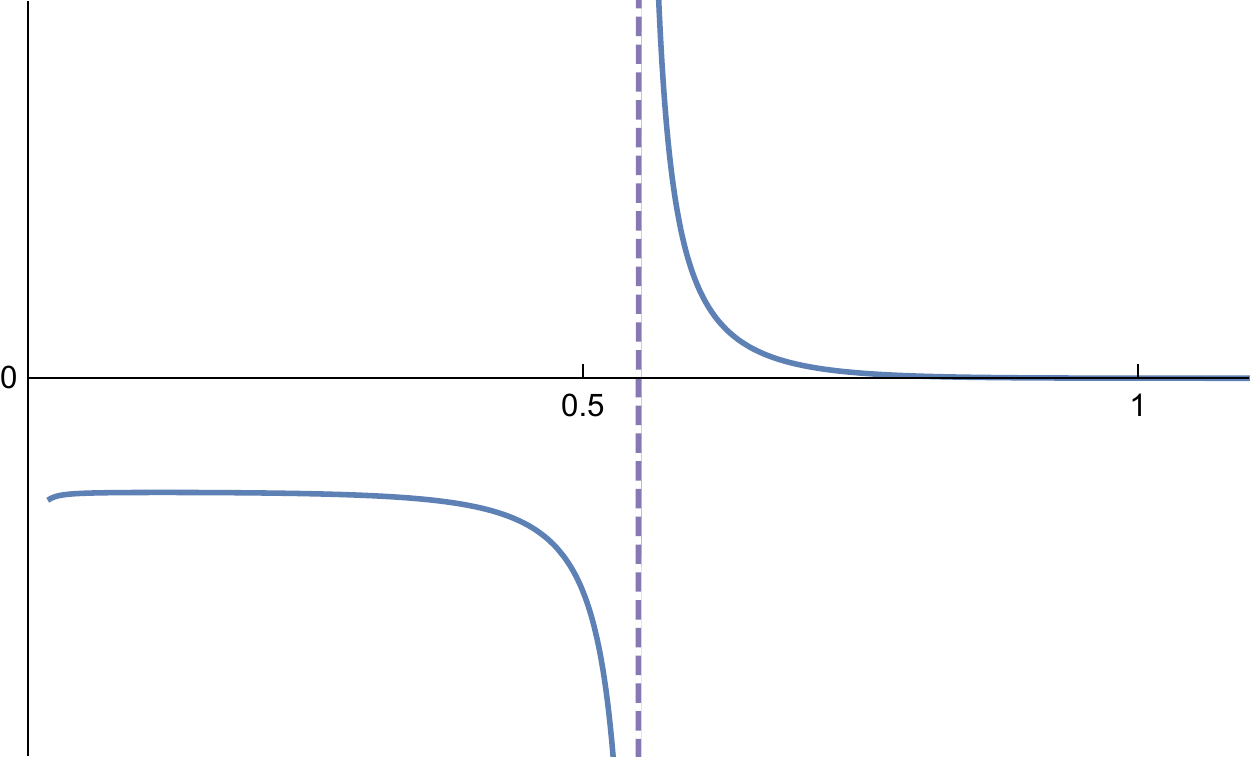}};
\node at (4,0) {\includegraphics[scale=0.4]{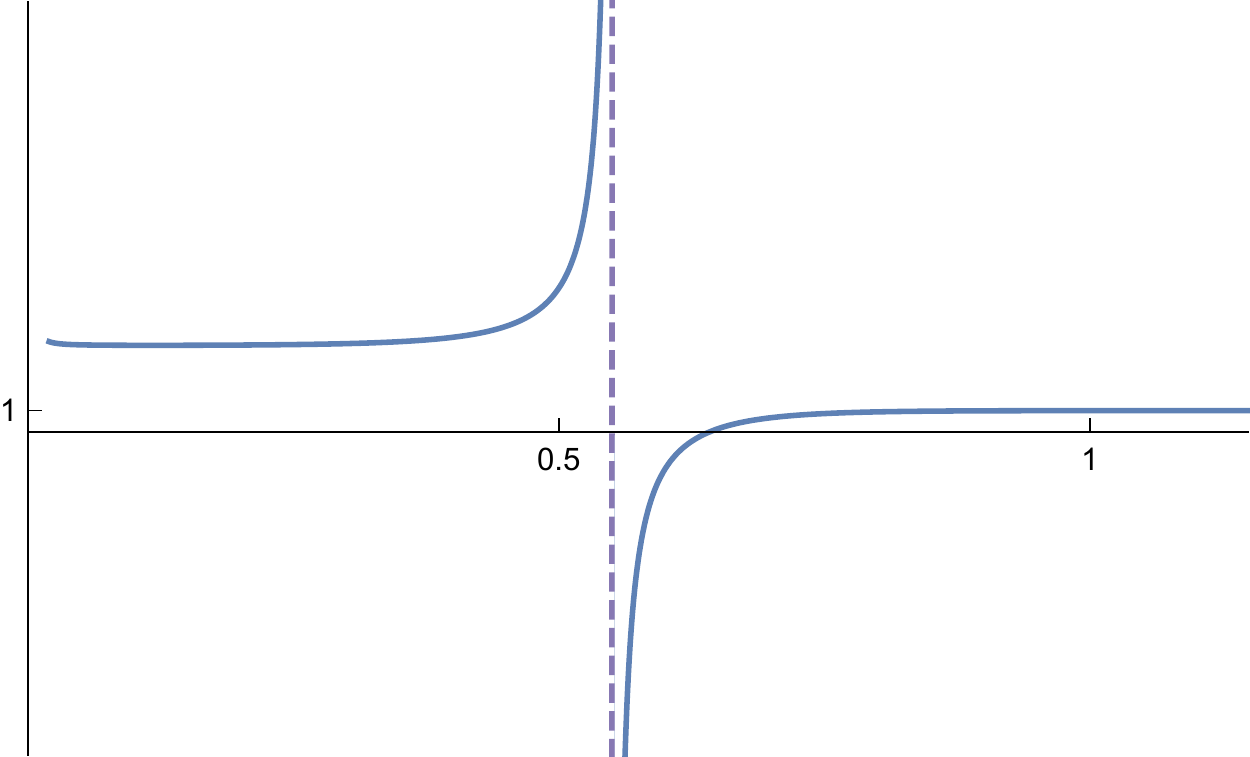}};
\node at (-10.25,1.8) {$m(r)/M$};
\node at (-4.4,1.75) {$1/\beta$};
\node at (1.65,1.75) {$4m(r)\kappa$};
\node at (-5.3,-1.35) {$r$};
\node at (0.7,0) {$r$};
\node at (6.7,-0.25) {$r$};
\fill [white](-7.95,-1.5) circle (.1cm);
\fill [white](-9.15,-1.5) circle (.1cm);
\fill [white](-6.75,-1.5) circle (.1cm);
\fill [white](-5.5,-1.5) circle (.1cm);
\fill [white](-10.45,1.4) circle (.1cm);
\fill [white](-10.45,0.0) circle (.1cm);
\node at (-7.95,-1.5) {\scriptsize{$1.0$}};
\node at (-9.15,-1.5) {\scriptsize{$0.5$}};
\node at (-6.75,-1.5) {\scriptsize{$1.5$}};
\node at (-10.6,1.4) {\scriptsize{$1.0$}};
\node at (-10.6,0.0) {\scriptsize{$0.5$}};
%%%%%%%
\fill [white](0.05,-.15) circle (.1cm);
\fill [white](-2.175,-.15) circle (.1cm);
\fill [white](-4.55,0) circle (.1cm);
\node at (0.05,-.15) {\scriptsize{$1.0$}};
\node at (-2.175,-.15) {\scriptsize{$0.5$}};
\node at (-4.55,0) {\scriptsize{$0$}};
%%%%%%%
\fill [white](5.95,-.35) circle (.1cm);
\fill [white](3.725,-.35) circle (.1cm);
\fill [white](1.45,-0.10) circle (.1cm);
\node at (5.95,-.35) {\scriptsize{$1.0$}};
\node at (3.725,-.35) {\scriptsize{$0.5$}};
\node at (1.45,-0.10) {\scriptsize{$1$}};
\end{tikzpicture}
\caption{Collapsing dust cloud: plots of the radial mass profile (left), the slope parameter $\beta$ (middle), the surface gravity $\kappa$ (right),  for $r_{*} = 1$, $q=1/20$, and $M = 1/100$. The region $r<r_{*}$ is the dense region. The change in the sign of $\beta$ indicates the change in of signature of $H$ from timelike to spacelike. The surface gravity saturates to $1/4M$ in the dilute region.}
\label{fig-col-beta3}
\end{figure*}
%\begin{figure}[h]
%\centering
%\includegraphics[scale=0.5]{collapse_mass3.pdf}
%\caption{Plot of $m(r)/M$ versus $r$ for $r_{*}=1$ and $q=1/20$. The region $r<r_{*}$ is the dense region.}
%\label{fig-col-mass2}
%\end{figure}

The metric describing the collapse is
\begin{equation}
ds^{2} = -d\tau^{2} + \frac{R'(\tau,r)^{2}}{1-2m(r)/r} dr^{2} + R^{2}\left(d\theta^{2} + \sin^{2}\theta d\phi^{2} \right) \, ,
\end{equation}
where $\tau$ is the proper time along the dust particles, and
\begin{equation}
m(r) = 4\pi \int_{0}^{r} dr' r'^{2} \rho(r') \, .
\end{equation}
The stress tensor is
\begin{equation}
T_{ab} = \frac{r^{2} \rho(r)}{R^{2}(\tau,r) R'(\tau,r)} (d\tau)_{a}(d\tau)_{b} \, .
\end{equation}
The future holographic screen satisfies~\cite{Booth-examples}
\begin{equation}
R(\tau,r) = 2 m(r). \label{eq-screen-collapse}
\end{equation}
The null normals such that $k^{a}l_{a} = -1$ and $\theta^{(l)} = -2/R$ are
\begin{eqnarray}
k^{a} & \cong & \sqrt{1 - \frac{2m(r)}{r}} \left(\frac{\partial}{\partial \tau}\right)^{a} + \frac{1-\frac{2m(r)}{r}}{R'(\tau,r)} \left(\frac{\partial}{\partial r}\right)^{a} \, ,\\
l^{a} & \cong & \frac{1}{2}\frac{1}{\sqrt{1 - \frac{2m(r)}{r}}} \left(\frac{\partial}{\partial \tau}\right)^{a} - \frac{1}{2R'(\tau,r)} \left(\frac{\partial}{\partial r}\right)^{a} \, ,
\end{eqnarray}
where $\cong$ means that we impose the constraint Eq.~(\ref{eq-screen-collapse}) while evaluating the right hand side. 
The slope $\beta$, and the surface gravity $\kappa$ are
\begin{eqnarray}
\beta & \cong & \frac{1}{2m'(r)} \frac{R' - m'(r)}{1 - 2m(r)/r} \, ,\\
\kappa & \cong & \frac{1}{2 m(r)} \, \frac{m'(r)}{R'(\tau,r)} \left( \frac{4\beta^{2}(1-2m(r)/r)^{2} - 1}{4\beta(1-2m(r)/r)} \right) \, .
\end{eqnarray}
As an example, we consider the `Fermi-Dirac distribution' for $\rho(r)$
\begin{equation}
\rho(r) = \frac{\text{M}}{-8\pi q^{3} \text{Li}_{3}\left(-e^{r_{*}/q}\right)} \frac{1}{\exp^{\left( \frac{r-r_{*}}{q}\right)}+1} \, ,
\end{equation}
where the overall normalization is such that $M$ is the ADM mass. This reduces to the standard Oppenheimer-Synder solution in the limit $q\rightarrow 0$. In the following, we fix $r_{*}=1$ and $q = 1/20$.

The profile of the mass is shown in Fig.~\ref{fig-col-beta3}. It shows that the most of the mass is in $r<r_{*}$. We call this region dense region. 

Fig.~\ref{fig-col-beta3} also shows the plots of $1/\beta$ and $4m(r)\kappa$. Note that $\beta$ is negative for small $r$ and then becomes positive. This implies a change in the signature of the screen. From the plot of $4m(r)\kappa$, we learn that the surface gravity quickly saturates to the Schwarzschild value in the near-vacuum region: 
\begin{equation}
\kappa \rightarrow \frac{1}{4M}~.
\end{equation}
%\begin{figure*}
%\begin{tikzpicture}
%\node at (-8,0) {\includegraphics[scale=0.4]{collapse_mass3.pdf}};
%\node at (-2,0) {\includegraphics[scale=0.4]{collapse_beta4.pdf}};
%\node at (4,0) {\includegraphics[scale=0.4]{collapse_kappa4.pdf}};
%\end{tikzpicture}
%\caption{Plot of $m(r)/M$ versus $r$ (left), plot of $1/\beta$ versus $r$ (middle) and plot of $4m(r)\kappa$ versus $r$ (right)  for $r_{*} = 1$, $q=1/20$, and $M = 1/100$. The region $r<r_{*}$ is the dense region. Change in the sign of $\beta$ implies the change in the signature of the screen. The surface gravity saturates to the Schwarzschild value outside the dense region.}
%\label{fig-col-beta3}
%\end{figure*}

\subsection{Charged black holes}
\label{sec-cv}

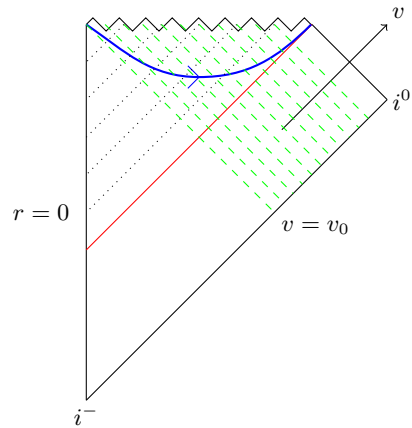
\begin{figure}[h]
\begin{tikzpicture}%[scale=0.8]
  \draw [black](0,-4) --(0,1);
	\draw [black](0,-4) --(4,0);
	\draw [black](4,0) --(3,1);
	\draw [->] (2.6,-0.4) --(4.0,1.0);
	\node at (4.15,1.15) {$v$};
	%\draw [fill=gray!60!white](0,-4) to [out=70,in=270] (1.5,1) --(0,1) --(0,-4);
	%\draw [fill=gray!20!white](0,-4) to [out=70,in=270] (1.5,1) --(3,1) --(4,0) --(0,-4);
	\draw [decorate,decoration=zigzag](0,1) --(3,1);
	\draw [red](3,1) --(0,-2);
	\draw [blue,thick](0,1) to [out=325,in=180] (1.5,0.3) to [out=0,in=225] (2.99,1);
	\draw [blue](1.35,.45) to (1.5,0.3) to (1.35,0.15);
	\node at (0,-4.2) {$i^{-}$};
	\node at (4.2,0) {$i^{0}$};
	\node at (-0.6,-1.5) {$r=0$};
	\draw [dotted](2.5,1) --(0,-1.5);
	\draw [dotted](2,1) --(0,-1);
	\draw [dotted](1.5,1) --(0,-.5);
	\draw [dotted](1,1) --(0,0);
	\draw [dotted](0.5,1) --(0,.5);
%	\draw [black](0,-9) --(0,5);
\draw [green,dashed](0,1) --(2.5,-1.5);
\draw [green,dashed](0.25,1) --(2.625,-1.375);
	\draw [green,dashed](.5,1) --(2.75,-1.25);
	\draw [green,dashed](.75,1) --(2.875,-1.125);
	\draw [green,dashed](1.0,1) --(3,-1.0);
	\draw [green,dashed](1.25,1) --(3.125,-0.875);
	\draw [green,dashed](1.5,1) --(3.25,-0.75);
	\draw [green,dashed](1.75,1) --(3.375,-0.625);
	\draw [green,dashed](2.0,1) --(3.5,-.5);
	\draw [green,dashed](2.25,1) --(3.625,-0.375);
	\draw [green,dashed](2.5,1) --(3.75,-.25);
	\node at (3.05,-1.7) {$v=v_{0}$};
 \end{tikzpicture}
\caption{The Penrose diagram for Vaidya solution. We show the uncharged case, $e(v) = 0$. The mass function is $m(v) = 0$ for $v<v_{0}$ and $\dot{m}(v) \ge 0$ for $v>v_{0}$. The green dashed lines are the ingoing null shells. The red line is the event horizon. The blue line is the future holographic screen constructed from future light-cones centered at $r=0$.}
\label{fig-VB}
\end{figure}

Next, we consider the Vaidya-Bonnor solution~\cite{Vaidya-1,Vaidya-2}, which describes the formation of a black hole by an arbitrary sequence of charged spherical null shells:
\begin{equation}
ds^{2} = -f\,dv^2 +2dv\,dr +r^{2}\,d\Omega^{2} \label{eq-VB-metric}
\end{equation}
where 
\begin{equation}
f(r,v) = 1-\frac{2m(v)}{r} + \frac{e^{2}(v)}{r^{2}}
\end{equation}
and $m(v) \ge |e(v)|$ are integrable differentiable functions with 
\begin{equation}
\dot m\equiv \frac{\partial m}{\partial v}>0~,~~ m(\infty)<\infty~.
\end{equation}
An example is shown in Fig.~\ref{fig-VB}. The stress tensor is%nonvanishing components of the stress tensor are
\begin{equation} 
T_{ab} = -\frac{1}{8\pi r} \dot{f}(r,v) (dv)_{a}(dv)_{b} + T_{ab}^{(EM)} \, ,
\end{equation} 
where 
\begin{equation} 
T_{ab}^{(EM)} = - \frac{1}{8\pi} \frac{e^{2}(v)}{r^{4}} \left( g_{ab} - 2 r^{2} s_{ab} \right) 
\end{equation} 
is the stress tensor of the point charge of magnitude $e(v)$, and $s_{ab}$ is the metric on the unit two-sphere. 

The holographic screen, $H$, consists of marginally trapped surfaces, with $k$ the future and outward directed null vector. The condition $\theta^{(k)}=0$ implies $r=R$ and $f(R,v)=0$, and thus 
\begin{equation}
R = m(v) + \sqrt{m^{2}(v) -e^{2}(v)}~.
\end{equation}
The null vectors normal to the leaves are found to be
\begin{eqnarray}
k^{a} & = & \left( \frac{\partial}{\partial v}\right)^{a} \, ,\\
l^{a} & = & - \left( \frac{\partial}{\partial r}\right)^{a} \, .
\end{eqnarray}
Their linear combinations tangent and normal to $H$, $h=-l+\beta k$, $n=l+\beta k$ are determined by
\begin{equation} 
\beta = \frac{dv}{dR} = \frac{R - m(v)}{R\dot{m}(v) - e(v)\dot{e}(v)}~. \label{eq-beta-vaidya}
\end{equation}
The acceleration is
\begin{equation}
\tilde\kappa = \frac{(R-m(v))^{2}}{R^{2}(R\dot{m}(v)-e(v)\dot{e}(v))}~;
\end{equation}
the surface gravity, $\tilde\kappa/\beta$, is
\begin{align}
\kappa = \frac{m(v)}{R^{2}} - \frac{e^{2}(v)}{R^{3}},
\end{align}

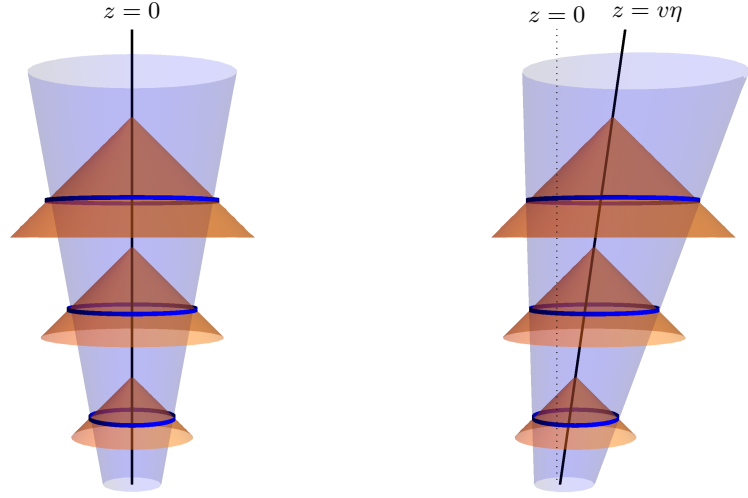
\begin{figure*}
\begin{tikzpicture}
%\node at (-8,0) {\includegraphics[scale=0.5]{collapse_mass2.pdf}};
\node at (-2,0) {\includegraphics[scale=0.5]{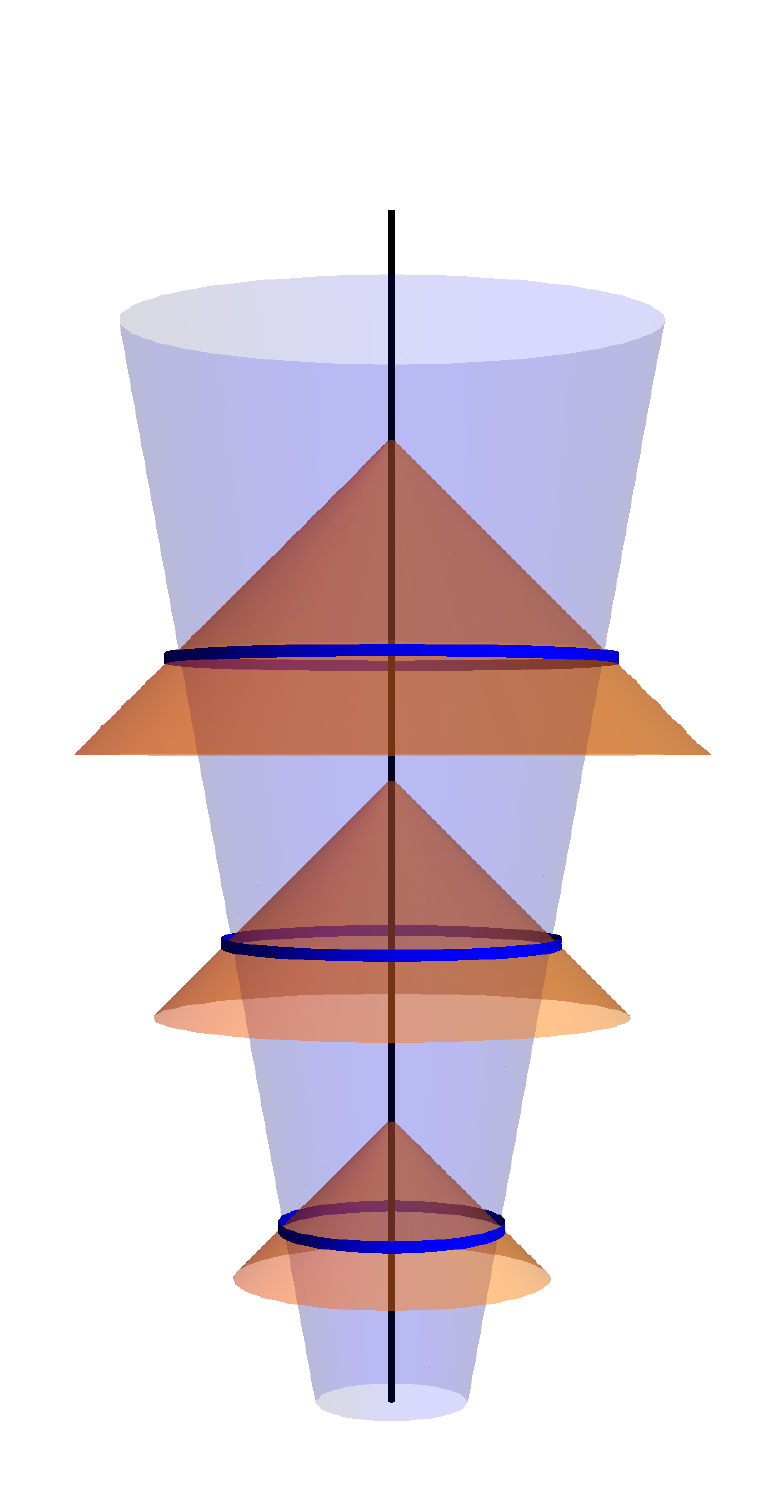}};
\node at (4,0) {\includegraphics[scale=0.5]{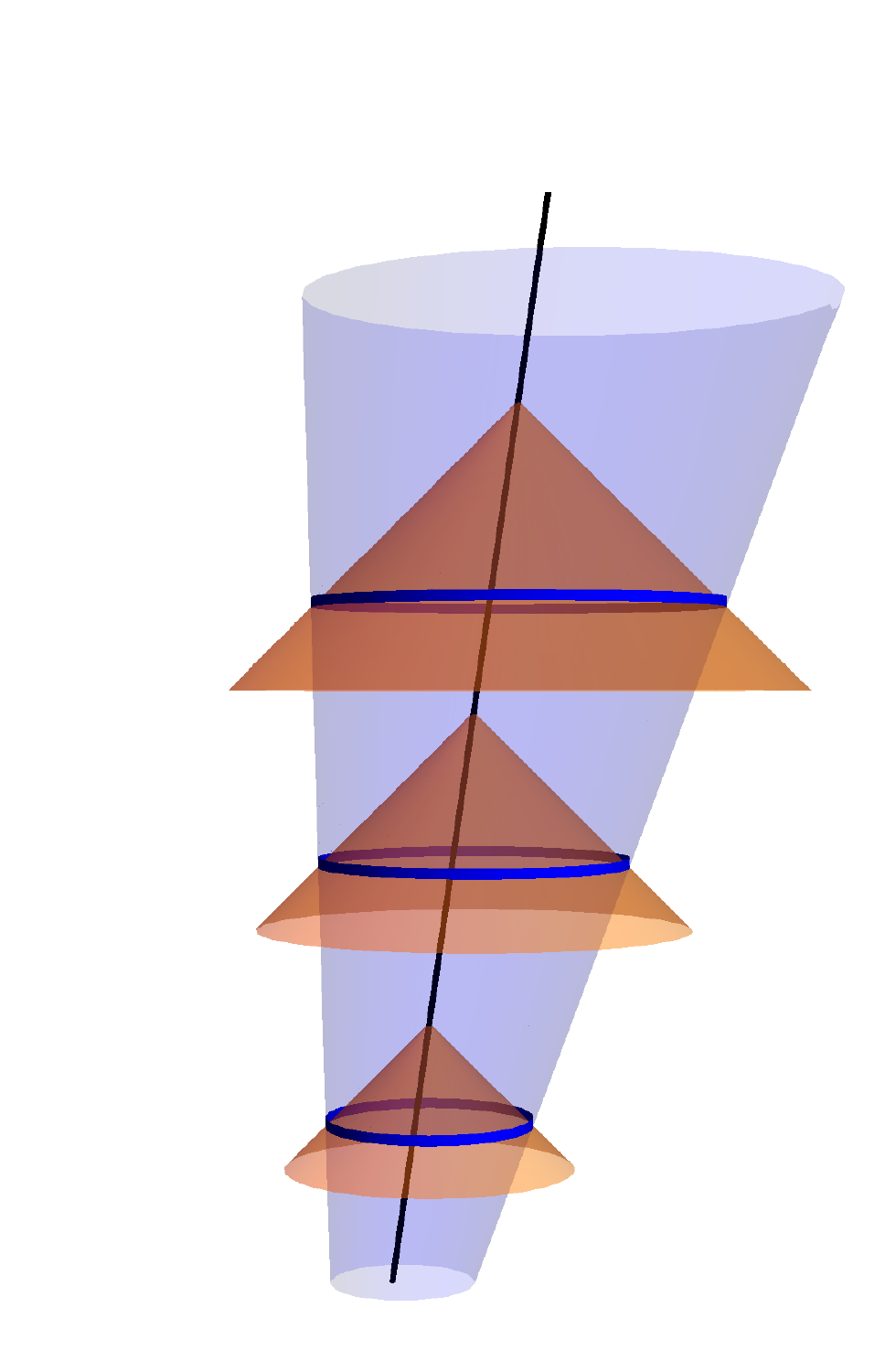}};
\node at (-2,3) {$z=0$};
\node at (4.85,2.95) {$z=v\eta$};
\node at (3.65,2.95) {$z=0$};
\draw [dotted](3.65,-3.25) -- (3.65,2.75);
\end{tikzpicture}
\caption{Two past holographic screens in the same expanding universe, associated with two different observers (thick black worldlines). {\em Left:} spherically symmetric screen constructed from a comoving observer at $r=0$ (see Sec.~\ref{sec-frw}). {\em Right:} screen constructed from the past light-cones of a non-comoving observer (Sec.~\ref{sec-nonsym}).}
\label{fig-frw-nonsym}
\end{figure*}
Note that this result has the same form as the surface gravity of a Reissner-Nordstrom black hole with mass $m$ and charge $e$. Moreover, let us define an electric potential in the usual way,
\begin{equation}
\Phi = \frac{e(v)}{R}~.
\end{equation}
Then Eq.~(\ref{conj}) takes a form similar to the first law of thermodynamics for a Reissner-Nordstrom black hole:
\begin{equation}
\frac{\kappa}{8\pi}\frac{dA}{dR} = \frac{dm}{dR} - \Phi\mbox{ }\frac{de}{dR}~,
\end{equation}
where we have used Eq.~(\ref{eq-beta-vaidya}) and the chain rule, $\beta \dot{f} = v'(R) \dot{f} = f'$.

\subsection{Nonspherical Screen in Cosmology}
\label{sec-nonsym}

We again consider the expanding universe of Sec.~\ref{sec-frw} and specialize to the matter dominated universe: $p = 0$. The metric is
\begin{equation}
ds^{2} = a^{2}(\eta)\left( -d\eta^{2} + dx^{2} + dy^{2} + dz^{2} \right) \, ,
\end{equation}
with 
\begin{equation}
a(\eta) = \eta^{2}/9~.
\end{equation}
We pick an observer whose worldline is given by 
\begin{equation}
z = v \eta~,
\end{equation}
for $-1<v<1$. To construct the past screen, we shoot past light cones from each point on this worldline, and find the cross-section of maximal area on each of these light cones; see Fig.~\ref{fig-frw-nonsym}. The collection of all these cross-section, that is the past screen, satisfies the condition
\begin{equation}
\left( z - \frac{3}{2}\eta\right)^{2} + x^{2} + y^{2} - \frac{\eta^{2}}{4} = 0. \label{ns-frw-screen}
\end{equation}

We choose to work in the coordinate system 
\begin{eqnarray}
z & = & \frac{3}{2}v\eta + r\cos\theta \, ,\\
x & = & r\sin\theta \cos\phi \, ,\\
y & = & r\sin\theta \sin\phi \, ,\\
\eta & = & \eta \, .
\end{eqnarray}
In these coordiante system, Eq.~(\ref{ns-frw-screen}) simplifies to
\begin{equation}
r - \frac{\eta}{2} = 0. 
\end{equation}
The area-radius of the leaf of the screen is
\begin{equation}
R = r a(\eta) = \frac{\eta^{3}}{18}.
\end{equation}
We pick an orthogonal basis for the one-forms tangent tn the leaf: %$q_{a}^{\, b}e^{(i)}_{b} = e^{(i)}_{a}$ for $i = \{1,2\}$ and $e^{(1)}\cdot e^{(2)} = 0$.
\begin{eqnarray}
\hat{e}^{(1)}_{a} & = & -\frac{3v\sin\theta}{\eta}(d\eta)_{a} + (d\theta)_{a} \, ,\\
\hat{e}^{(2)}_{a} & = & (d\phi)_{a} \, ,
\end{eqnarray}
where $q_{a}^{~ b} \hat{e}^{(i)}_{b} = \hat{e}^{(i)}_{a}$ for $i = \{ 1,2 \}$, and $\hat{e}^{(1)}\cdot \hat{e}^{(2)} = 0$. 
Similarly, we pick an orthogonal basis for vectors normal to the leaf:
\begin{eqnarray}
\chi_{(1)}^{a} & = & \left( \frac{\partial}{\partial\eta}\right)^{a} - \frac{3 v \cos\theta}{2}\left( \frac{\partial}{\partial r}\right)^{a} + \frac{3v\sin\theta}{\eta}\left( \frac{\partial}{\partial\theta}\right)^{a} \, ,\\
\chi_{(2)}^{a} & = & \left( \frac{\partial}{\partial r}\right)^{a} \, .
\end{eqnarray}
where $q^{a}_{~ b} \chi_{(i)}^{b} = 0$ for $i = \{ 1,2 \}$, and $\chi_{(1)}\cdot \chi_{(2)} = 0$.  The null vectors normal to the leaf, normalized such that $k^{a}l_{a}=-1$ and $\theta^{(l)}=2/R$, are
\begin{eqnarray}
k^{a} & = & \frac{1}{a(\eta)}  \left( \chi_{(1)}^{a} - \chi_{(2)}^{a} \right) \, ,\\
l^{a} & = & \frac{1}{2a(\eta)}   \left( \chi_{(1)}^{a} + \chi_{(2)}^{a}  \right) \, .
\end{eqnarray}
The tangent and normal vectors to the screen are $h^{a} = \alpha l^{a} + \beta k^{a}$ and $n^{a} = -\alpha l^{a} + \beta k^{a}$, where 
\begin{eqnarray}
\alpha & = & 1 + v \cos\theta \, , \label{eq-alpha-nonsym-frw}\\
\beta & = & \frac{1}{6} \left( 1 - 3 v\cos\theta \right) \, .
\end{eqnarray}
The normal one-form and the acceleration are 
\begin{eqnarray}
%\theta^{(l)} & = & \frac{2}{R} \, ,\\
\Omega_{a} & = & 0 \, ,\\
\widehat{\Omega}_{a} & = & - \frac{2 v \sin\theta}{3} \,  \hat{e}^{(1)}_{a} \, ,\\
\tilde{\kappa} & = & -\frac{1}{3R} \left( 1+3v\cos\theta \right) \, .
\end{eqnarray}

One can easily combine this construction with that of Sec.~\ref{sec-frw}. For example, we can use the worldline $r=0$ up to some conformal time $\eta_*$ to construct a portion of the screen which is centered at $r=0$. We can then consider continuing this worldline to that of a moving observer, by substituting $\eta\to\eta-\eta_*$ in Eq.~(\ref{ns-frw-screen}) and below.  This corresponds to choosing $\alpha$ as in Eq.~(\ref{eq-alpha-nonsym-frw}), instead of $\alpha = 1$, for $\eta>\eta_*$. We thus obtain a nonspherical screen (above $\eta_*$), corresponding to the fact that the observer's worldline (and the associated null foliation) violates the spherical symmetry, above $\eta_*$. This illustrates how the observer-dependence of $H$ is captured by a choice of $\alpha$, as advertised in Sec.~\ref{sec-dod2}.

\label{sec-screens}
\vskip .3cm
\indent {\bf Acknowledgments} 
It is a pleasure to thank N.~Engelhardt, B.~Krishnan, N.~Obers, and M.~Rangamani for discussions. This work was supported in part by the Berkeley Center for Theoretical Physics, by the National Science Foundation (award numbers 1521446 and 1316783), by FQXi, and by the US Department of Energy under Contract DE-AC02-05CH11231..

%%%%%%%%%%%%%%%
\appendix	

\section{Parallel Transport of Null Vectors}%Derivation of Eqs.~(\ref{eq-kder})-(\ref{eq-lder})}
\label{sec-partran}

Here we present a derivation of Eq.~(\ref{eq-kder}); Eq.~(\ref{eq-lder}) can be derived similarly. We start with the ansatz
\begin{equation}
h^{b}\nabla_{a}k_{a} = A k_{a} + B l_{a} + C_{a} \, , \label{eq-pt-ans}
\end{equation}
where $C_{a}$ is the projection of $h^{b}\nabla_{b}k_{a}$ onto the leaf.
The constant $A$ is 
\begin{equation}
A = -l^{a}h^{b}\nabla_{b}k_{a} = \tilde{\kappa} \, ,
\end{equation}
where we have used Eq.~(\ref{eq-kappa}). The constant $B$ is 
\begin{equation}
B =-k^{a}h^{b}\nabla_{b}k_{a} = 0 \, .
\end{equation}
To determine $C_{a}$, we consider an arbitrary vector field, $\phi^{a}$, tangent to the leaf, and contract it to our ansatz%with both sides of Eq.~(\ref{eq-pt-ans}) 
\begin{eqnarray}
\phi^{a}C_{a} & = & \phi^{a}h^{b}\nabla_{b}k_{a} \, , \\
%& = & -k_{a} \mathcal{L}_{h}\phi^{a} - k_{a}\phi^{b}\nabla_{b}h^{a} \, ,\\
& = & -k_{a} \mathcal{L}_{h}\phi^{a} + \phi^{a} \left( D_{a}\alpha - \alpha\Omega_{a}\right) \, , \label{eq-phiC}
\end{eqnarray}
where we have used Eq.~(\ref{eq-omega-one-form}). A consequence of the normalization Eq.~(\ref{eq-hR}) is that for every vector field $\phi^{a}$ tangent to the leaf, $\mathcal{L}_{h}\phi^{a}$ is also tangent to the leaf \cite{Gourgoulhon-TH}
\begin{equation}
q^{a}_{~ b}\phi^{b} = \phi^{a} ~ \Rightarrow ~ q^{a}_{~ b}\mathcal{L}_{h}\phi^{b} = \mathcal{L}_{h}\phi^{a} \, .
\end{equation}
This with the fact that $\phi^{a}$ is arbitrary implies
\begin{equation}
C_{a} = D_{a}\alpha - \alpha\Omega_{a} \, .
\end{equation}
Eq.~(\ref{eq-pt-ans}) thus reduces to the desired result% Eq.~(\ref{eq-kder})
\begin{equation}
h^{b}\nabla_{b}k_{a} =  \tilde\kappa k_{a} + D_{a}\alpha - \alpha\Omega_{a} \, .
\end{equation}
	
\begin{widetext}
\section{Cross-focusing Equations} \label{cross_focussing_derivation}

Here we  derive  the cross-focusing equations~\cite{Hayward-TH,Hayward-law-2}
\begin{eqnarray}
	l^{a}\nabla_{a}\theta^{(k)} & = & -\theta^{(l)}\theta^{(k)} - \frac{1}{2}\mathcal{R} + \Omega_{a}\Omega^{a} - D_{a}\Omega^{a} -\frac{2}{\alpha}\Omega^{a}D_{a}\alpha + \frac{1}{\alpha}D^{a}D_{a}\alpha + 8\pi T_{ab}k^{a}l^{b} \, ,\label{eq-cross-focusing-one}\\
	k^{a}\nabla_{a}\theta^{(l)} + \kappa\theta^{(l)} & = & -\theta^{(l)}\theta^{(k)} - \frac{1}{2}\mathcal{R} + \Omega_{a}\Omega^{a} + D_{a}\Omega^{a} + \frac{2}{\beta}\Omega^{a}D_{a}\beta + \frac{1}{\beta}D^{a}D_{a}\beta + 8\pi T_{ab}k^{a}l^{b} \, ,\label{eq-cross-focusing-two}
\end{eqnarray}
\end{widetext}
	which will be useful in the derivation of the screen equations in Appendix~\ref{der_screen_eqs}. Note that these equations (unlike the screen equations) are highly sensitive to how we extend the null vectors, $k^{a}$ and $l^{a}$, into a neighborhood of the holographic screen. We do this by demanding 
	\begin{eqnarray}
	l^{b}\nabla_{b}l^{a} & = & 0 \, ,\label{eq-l-off-H}\\
	k^{b}\nabla_{b}k^{a} & = & \kappa k^{a} \, ,\\
	l^{a}k_{a} & = & -1 \, .
	\end{eqnarray}
	With these extension, Eq.~(\ref{eq-kder}) reduces to 
	\begin{equation}
	l^{b}\nabla_{b}k_{a} = \frac{1}{\alpha}D_{a}\alpha - \Omega_{a} \, .
	\end{equation}
	We decompose $\nabla_{a}l_{b}$ and $\nabla_{a}k_{b}$ as
	\begin{eqnarray}
	%\nabla_{a}l_{b} & = & g_{a}^{~ c}g_{b}^{~ d}\nabla_{c}l_{d} \, ,\\
	%& = & \left(q_{a}^{~ c} -l_{a}k^{c} - k_{a}l^{c} \right)\left(q_{b}^{~ d} -l_{b}k^{d} - k_{b}l^{d} \right)\nabla_{c}l_{d} \, ,\\
	\nabla_{a}l_{b} & = & B^{(l)}_{ab} + \kappa l_{a}l_{b} - l_{a}\Omega_{b} - l_{b}\Omega_{a} - \frac{1}{\beta}l_{a}D_{b}\beta \, ,\label{dl}\\
	\nabla_{a}k_{b} & = & B^{(k)}_{ab} -\kappa l_{a}k_{b} + k_{a}\Omega_{b} + k_{b}\Omega_{a} - \frac{1}{\alpha}k_{a}D_{b}\alpha \, .\label{dk}
	\end{eqnarray}
	Now using $\theta^{(k)} = q^{ab}\nabla_{a}k_{b}$, we get
	\begin{widetext}
	\begin{eqnarray}
	l^{a}\nabla_{a}\theta^{(k)} & = & l^{a}\nabla_{a}\left( q^{bc}\nabla_{b}k_{c} \right) \, ,\\
	& = & q^{bc}\nabla_{b}\left( l^{a}\nabla_{a}k_{c}\right) - q^{bc}\left(\nabla_{b}l^{a}\right)\left(\nabla_{a}k_{c}\right) + \left(l^{a}\nabla_{a}q^{bc}\right)\left(\nabla_{b}k_{c}\right) + R_{abcd}l^{a}q^{bc}k^{d} \, ,\\
	& = & \Omega_{a}\Omega^{a} - D_{a}\Omega^{a} - B^{(l)}_{ab}B_{(k)}^{ab} -\frac{2}{\alpha}\Omega^{a}D_{a}\alpha + \frac{1}{\alpha}D^{a}D_{a}\alpha +  R_{abcd}l^{a}q^{bc}k^{d} \, .
	\end{eqnarray}
	%\end{widetext}
	With the help of the Gauss-Codazzi equation for codimension-$2$ spatial surfaces~\cite{Hayward:1993ph},
	\begin{equation}
	\frac{1}{2}\mathcal{R} + \theta^{(l)}\theta^{(k)} - B^{(l)}_{ab}B_{(k)}^{ab} = \frac{1}{2}R_{abcd}q^{ac}q^{bd} \, ,
	\end{equation}
	we get
	%\begin{widetext}
	\begin{eqnarray}
	l^{a}\nabla_{a}\theta^{(k)} & = & -\theta^{(l)}\theta^{(k)} - \frac{1}{2}\mathcal{R} + \Omega_{a}\Omega^{a} - D_{a}\Omega^{a} -\frac{2}{\alpha}\Omega^{a}D_{a}\alpha + \frac{1}{\alpha}D^{a}D_{a}\alpha + \frac{1}{2}R_{ab}q^{ab} \, .
	\end{eqnarray}
	%\end{widetext}
	To get Eq.~(\ref{eq-cross-focusing-one}), we use the Einstein equations
	\begin{equation}
	\frac{1}{2}R_{ab}q^{ab} = \left(R_{ab} - \frac{1}{2}R g_{ab}\right)k^{a}l^{b} = 8\pi T_{ab}k^{a}l^{b} \, .
	\end{equation}
	Eq.~(\ref{eq-cross-focusing-two}) can be derived in a similar fashion.
%	\end{widetext}

%%%%%%%%%%%%%%%%
\section{Derivation of Screen Equations} \label{der_screen_eqs}

Here we present a derivation of the nontrivial screen equations, (\ref{eq-kc})-(\ref{eq-bc}). Eq.~(\ref{eq-bc}) was derived for the dynamical horizons in Refs.~\cite{AshKri02,AshKri03}. The derivation made use of the Gauss Codazzi constraint equations which relate the extrinsic curvature of a hypersurface with the Ricci tensor of the background spacetime. This can only be done for a hypersurface with definite signature. Hence, this method does not obviously apply to a holographic screen. In Sec.~(\ref{app-con-eq}), we will present a derivation of Eq.~(\ref{eq-bc}) in a way that makes it clear that the signature of $H$ (and indeed, anything but the $2D$ leaf) is irrelevant. Eq.~(\ref{eq-kc}) was derived in Ref.~\cite{Gourgoulhon-DH}; in Sec.~(\ref{app-sc-eq}) , we present a simplified derivation. In Sec.~(\ref{app-vec-eq}), we derive Eq.~(\ref{eq-omegadyn}), following Ref.~\cite{Gourgoulhon-TH}.

\subsection{$T_{ab}n^{a}k^{b}$ Equation} \label{app-con-eq}
We begin by deriving Eq.~(\ref{eq-bc}). On the holographic screen, $h^{a}\nabla_{a}\theta^{(k)} = 0$. Expanding this equation, we get
\begin{equation}
0 = \alpha l^{a}\nabla_{a}\theta^{(k)} + \beta k^{a}\nabla_{a}\theta^{(k)} \, . \label{eq-screen-cons-int}
\end{equation}
Replacing the first term on the right hand side with the cross-focusing Eq.~(\ref{eq-cross-focusing-one}), and the second term with Raychaudhuri's equation,
%\begin{widetext}
\begin{equation}
k^{a}\nabla_{a}\theta^{(k)} = \kappa\theta^{(k)} - \frac{1}{2}\theta^{(k)2}-\sigma^{(k)}_{ab}\sigma_{(k)}^{ab} - 8\pi T_{ab}k^{a}k^{b} \, ,
\end{equation}
we find
%\begin{widetext}
\begin{equation}
-\frac{\alpha}{2} \mathcal{R} +\alpha \Omega_a\Omega^a  -\alpha D_{a}\Omega^{a} -2\Omega^aD_a\alpha+D_aD^a\alpha  = 8\pi T_{ab} n^a k^b  +\beta\sigma^{(k)}_{ab}\sigma_{(k)}^{ab}~. \label{eq-sc-con-appendix}
\end{equation}
%\end{widetext}

\subsection{$T_{ab}n^{a}h^{b}$ Equation} \label{app-sc-eq}
Next, we derive Eq.~(\ref{eq-kc}). By Eq.~(\ref{eq-hbeta}), 
\begin{equation}
\alpha \widehat{\mathcal{L}}_{h}\theta^{(l)} = \alpha^{2} l^{a}\nabla_{a}\theta^{(l)} + \alpha\beta k^{a}\nabla_{a}\theta^{(l)}~.
\end{equation}
We replace the second term on the right hand side with the cross-focusing Eq.~(\ref{eq-cross-focusing-two}) and the first term with Raychaudhuri's equation
\begin{equation}
l^{a}\nabla_{a}\theta^{(l)} = -\frac{1}{2}\theta^{(l)2} - \sigma^{(l)}_{ab}\sigma_{(l)}^{ab} - 8\pi T_{ab}l^{a}l^{b} \, .
\end{equation}
As a result, we get%to get
%\begin{widetext}
\begin{equation}
\alpha (\widehat{\mathcal{L}}_{h} + \tilde{\kappa})\theta^{(l)} = -\alpha^{2}B^{(l)}_{ab}B^{ab}_{(l)} -\frac{1}{2}\alpha\beta\mathcal{R} + \alpha\beta\Omega^{a}\Omega_{a} +\alpha\beta D_{a}\Omega^{a} + 2\alpha \Omega^{a}D_{a}\beta + \alpha D^{a}D_{a}\beta +  8 \pi \alpha T_{ab}n^{a}l^{b}~.
\end{equation}
We eliminate $\mathcal{R}$ from this equation by using Eq.~(\ref{eq-sc-con-appendix})
\begin{equation}
\alpha (\widehat{\mathcal{L}}_{h} + \tilde{\kappa})\theta^{(l)} =  D_{a}(2\alpha\beta\Omega^{a} - \beta D^{a}\alpha +\alpha D^{a}\beta) + B^{(h)}_{ab}B^{ab}_{(n)} + 8 \pi T_{ab}n^{a}h^{b}~.
\end{equation}
Rearranging the terms and using Eq.~(\ref{eq-exp-tilde-omega}) lead to Eq.~(\ref{eq-kc}).
%\end{widetext}

\subsection{$T_{ab} n^a q^b_{~c}$ Equations} \label{app-vec-eq}

Here we closely follow the derivation by Gourgoulhon~\cite{Gourgoulhon-TH}. We start with the identity
\begin{equation}
R_{ab}n^{a}q^{b}_{~ c} = q_{c}^{~ b}\left( \nabla_{a}\nabla_{b} - \nabla_{b}\nabla_{a}\right)n^{a} \, . \label{eq-sc-vec-int}
\end{equation}
%Using $n^{a} = -\alpha l^{a} + \beta k^{a}$, and Eqs.~(\ref{dl})-(\ref{dk}), we get
%\begin{widetext}
Using $n^{a} = -\alpha l^{a} + \beta k^{a}$, and Eqs.~(\ref{dl})-(\ref{dk}), we get
\begin{equation}
\nabla_{a}n_{b} = B_{ab}^{(n)} -\kappa l_{a}h_{b} + h_{a}\Omega_{b} + h_{b}\Omega_{a} + \frac{\alpha}{\beta}l_{a}D_{b}\beta - \frac{\beta}{\alpha}k_{a}D_{b}\alpha - l_{b}\nabla_{a}\alpha + k_{b}\nabla_{a}\beta~.
\end{equation}
The first term on the right hand side of Eq.~(\ref{eq-sc-vec-int}) becomes
\begin{equation}
q_{c}^{~ b}\nabla_{a}\nabla_{b}n^{a} = q_{c}^{~ b}h^{a}\nabla_{a}\Omega_{b} + \Omega^{a}B^{(h)}_{ac} + \theta^{(h)}\Omega_{c} - \theta^{(l)}D_{c}\alpha + q_{cb}\nabla_{a}B_{(n)}^{ab} -  B^{(n)}_{ca} \left( \frac{1}{\beta}D^{a}\beta + \frac{1}{\alpha}D^{a}\alpha\right) -D_{c}\left( l^{a}\nabla_{a}\alpha - k^{a}\nabla_{a}\beta\right) \, ,
\end{equation}  
where we have repeatedly use Eqs.~(\ref{eq-l-off-H})-(\ref{dk}). Similarly, the second term becomes
\begin{equation}
q_{c}^{~ b}\nabla_{b}\nabla_{a}n^{a} = D_{c}(\tilde{\kappa}) - \theta^{(l)}D_{c}\alpha - \alpha D_{c}\theta^{(l)} -D_{c}\left( l^{a}\nabla_{a}\alpha - k^{a}\nabla_{a}\beta\right) \, .
\end{equation}
Combining these two results, we get
\begin{equation}
R_{ab}n^{a}q^{b}_{~ c} = q_{c}^{~ b}h^{a}\nabla_{a}\Omega_{b} + \Omega^{a}B^{(h)}_{ac} + \theta^{(h)}\Omega_{c} - D_{c}\tilde{\kappa} +\alpha D_{c}\theta^{(l)} + q_{cb}\nabla_{a}B_{(n)}^{ab} -  B^{(n)}_{ca} \left( \frac{1}{\beta}D^{a}\beta + \frac{1}{\alpha}D^{a}\alpha\right) \, . \label{eq-sc-vec-int-two}
\end{equation}  
By making use of
\begin{equation}
D_{a}B^{(n)a}_{c} = q_{c}^{~ b}\nabla_{a}B^{(n)a}_{b} - B^{(n)b}_{c}\left(\frac{1}{\beta}D_{b}\beta + \frac{1}{\alpha}D_{b}\alpha\right) \, ,
\end{equation}
and 
\begin{equation}
\widehat{\mathcal{L}}_{h}\Omega_{c} = q_{c}^{~ b}h^{a}\nabla_{a}\Omega_{b} + \Omega^{a}B^{(h)}_{ac} \, ,
\end{equation}
Eq.~(\ref{eq-sc-vec-int-two}) reduces to
\begin{equation}
R_{ab}n^{a}q^{b}_{~ c} = \widehat{\mathcal{L}}_{h}\Omega_{c} + \theta^{(h)}\Omega_{c} - D_{c}\tilde{\kappa} +\alpha D_{c}\theta^{(l)} + D_{a}B^{(n)a}_{c} \, .
\end{equation}
Finally we use Einstein's equation to obtain Eq.~(\ref{eq-omegadyn}),
\begin{equation}
(\widehat{\mathcal{L}}_{h}+\theta^{(h)})\Omega_{c}- D_{c}\tilde\kappa + \alpha D_c\theta^{(l)} =  8\pi T_{ab}n^{a}q^{b}_{c} - D_{a}B_{c}^{(n)a}~.
\end{equation}
\end{widetext}

\bibliographystyle{utcaps}
\bibliography{all}
\end{document}